\documentclass[%
 reprint,
nofootinbib,
 amsmath,amssymb,
 aps,
pra,
floatfix,
nofootinbib,
noeprint,
nolongbibliography
]{revtex4-2}

\usepackage{placeins}
\usepackage{tabularx}
\usepackage{comment}
\usepackage{graphicx}
\usepackage{dcolumn}
\usepackage{bm}
\usepackage{color}
\usepackage{xfrac}
\usepackage{sidecap}

\newcommand{\ket}{\rangle}
\newcommand{\bra}{\langle}

\newcommand{\E}{\mathcal{E}}
\newcommand{\B}{\mathcal{B}}
\newcommand{\M}{\mathcal{M}}
\newcommand{\Eeff}{\mathcal{E}_{ef\!f}}

\usepackage{rotating}

\usepackage[dvipsnames]{xcolor}

\usepackage{threeparttable}

\usepackage{here}

\begin{document}

\title{Engineered Molecular Clock Transitions for Symmetry Violation Searches}

\author{Yuiki Takahashi}
\email{yuiki@caltech.edu}
\author{Harish D. Ramachandran}
\thanks{Present address: Department of Physics, The Johns Hopkins University, Baltimore, MD 21218 USA}
\author{Arian Jadbabaie}
\thanks{Present address: Department of Physics, Massachusetts Institute of Technology, Cambridge, MA 02139 USA}
\author{Yi Zeng}
\thanks{Present address: Institut f\"{u}r Experimentalphysik, Universit\"{a}t Innsbruck, A-6020 Innsbruck, Austria}
\author{Chi Zhang}
\author{Nicholas R. Hutzler}
\affiliation{Division of Physics, Mathematics, and Astronomy, California Institute of Technology, Pasadena, California 91125 USA}

\date{\today}

\begin{abstract}

Heavy polar molecules are sensitive probes of physics Beyond the Standard Model. However, uncontrolled external electromagnetic fields pose challenges to achieving precise and accurate measurements. 
Minimizing susceptibility to these fields is therefore critical and has played an important role in all precision experiments of this type. Here we devise and demonstrate clock transitions engineered to realize robust symmetry violation searches in the polyatomic molecule YbOH. Sensitivities to external fields can be suppressed by orders-of-magnitude while preserving high sensitivity to the electron electric dipole moment (eEDM). We perform Ramsey measurements on these clock transitions and observe suppression of electric and magnetic sensitivities by at least a factor of 700 and 200, respectively, and demonstrate the robustness of their spin coherence against large electromagnetic field fluctuations. We further identify and employ selected quantum states to make sensitive measurements of external magnetic and electric fields, another critical feature for highly accurate measurements. This approach of molecular engineering is broadly applicable to diverse molecular species and states, including those with complex nuclei and those that are compatible with state-of-the-art cooling and trapping techniques, thereby offering the potential to significantly improve experimental sensitivity to a wide range of New Physics while expanding the chemical design space for molecular quantum science.

\end{abstract}

\maketitle

Heavy polar molecules are powerful quantum sensors for probing physics Beyond the Standard Model (BSM)~\cite{Hinds1997, Safronova2018,Hutzler2020Review,DeMille2024}. A particularly promising route is searching for electromagnetic moments whose nonzero value would indicate time (T) and charge-parity (CP) symmetry violation, such as the electron’s electric dipole moment (eEDM)~\cite{Safronova2018}. The most sensitive eEDM searches, performed with HfF$^+$~\cite{Roussy2023_JILA_EDM_GEN2} and ThO~\cite{ACME2018}, probe new CP-violating physics at energies of around tens of TeV. These experiments achieve high sensitivity by leveraging the enhancement of CP-violating effects due to the large internal electric field in heavy polar molecules combined with two key advantages of a $^3\Delta_1$ molecular state. First, these states have a reduced magnetic moment of $\approx0.01\mu_B$, where $\mu_B$ is the Bohr magneton, which minimizes sensitivity to unwanted magnetic field effects -- a pernicious challenge in EDM experiments with atoms~\cite{Regan2002} and neutrons~\cite{Abel2020nEDM}.  Second, these states can be polarized in small electric fields $\lesssim 10$~V/cm and give rise to ``internal co-magnetometers,'' which enable reversal of the eEDM-induced energy shifts independently of the laboratory fields~\cite{Roussy2023_JILA_EDM_GEN2,ACME2018}. 
Nevertheless, residual sensitivities to uncontrolled fields remain a critical challenge, giving rise to systematic shifts that can mimic an eEDM signal.  Consequently, both experiments require careful measures to correct and reject these effects, and rely very heavily on the intrinsic reduced sensitivity and co-magnetometry offered by the $^3\Delta_1$ state.

Combining modern quantum tools with molecular CP-violation searches offers the potential to improve sensitivities to new physics by orders of magnitude, reaching to PeV energy scales~\cite{Alarcon2022Snowmass}.
However, molecules with high CP-violation sensitivity typically do not feature $^3\Delta_1$ states, and $^3\Delta_1$ structures are not amenable to laser cooling or ultracold assembly -- mainstay techniques that could dramatically enhance experimental sensitivity, especially for neutral species. Laser-coolable molecules typically feature a $^2\Sigma$ structure arising from a single, metal-centered valence electron~\cite{Isaev2016Poly,Kozyryev2016Poly,Fitch2021Review}, and as a result have magnetic moments on the order of $\sim\mu_B$, which is roughly 100 times larger than that of the $^3\Delta_1$ state.
This makes it challenging to utilize laser-coolable molecules for CP-violation searches given that stray electromagnetic fields remain a primary source of systematic errors~\cite{collings2025}.  
Additionally, molecules used to search for nuclear CP-violation typically have very complicated hyperfine structure~\cite{Grasdijk2021TlF,Takahashi2023magicPRL}, making demonstrated methods largely inapplicable.
While there are methods to overcome some of these challenges~\cite{verma_electron_2020, Ho_YbF_MQM_2023, Anderegg2023CaOHSpin}, they require difficult measurement schemes, lack co-magnetometry, or are confined to specific species and experimental techniques.  It is therefore useful to find a simple approach which offers high sensitivity to CP-violation and strong robustness against systematic errors without relying on a specific molecular structure.

\begin{figure*}[]
    \includegraphics[width=1\textwidth]{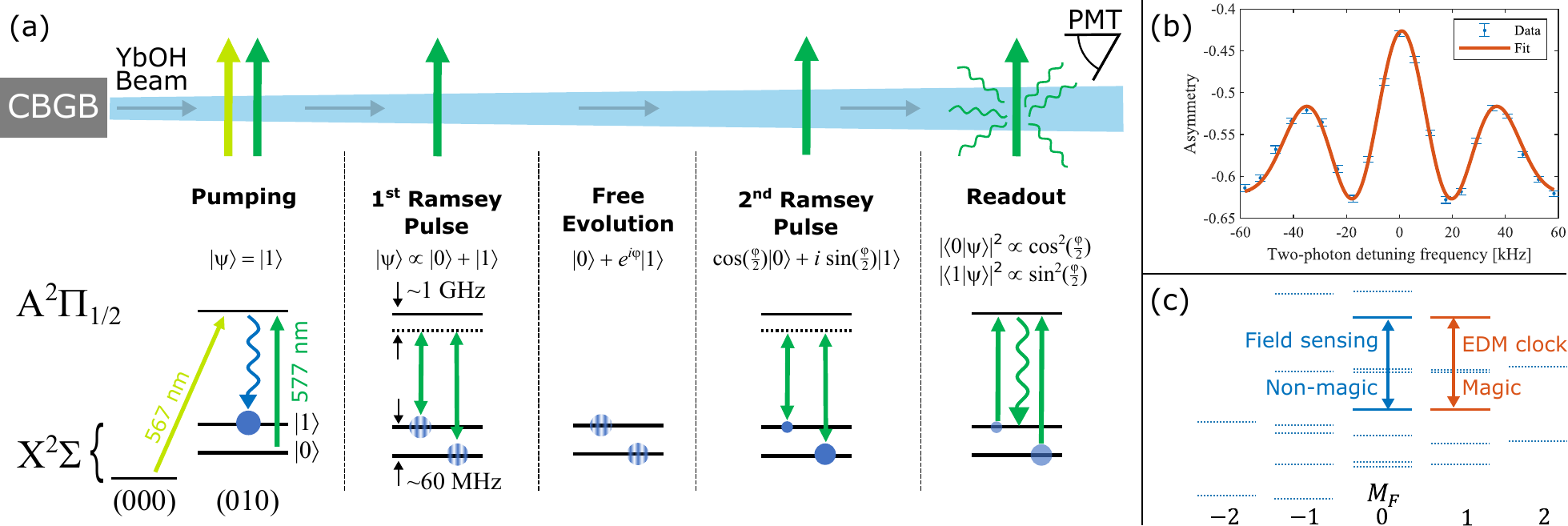}
\caption{(a) A schematic representation of the experiment.  A beam of YbOH molecules is directed through a series of lasers.  The first lasers optically pump from the ground, non-vibrating $(000)$ state to populate the $|1\ket$ level in the $(010)$ bending mode ``science state.''  Next, a two-photon pulse creates a coherent superposition of $|0\ket$ and $|1\ket$.  During spin precession, a phase $\varphi$ is accumulated. A second two-photon pulse then maps the relative phase onto populations of $|0\ket$ and $|1\ket$. (The second Ramsey pulse is applied by retro-reflecting the laser for the first Ramsey pulse.) Finally, state populations are determined by state-selective laser-induced fluorescence detected on a photomultiplier tube (PMT). (b) An example Ramsey fringe taken with $\E=10.93$ V/cm and $\B=0.30$ G as a function of two-photon frequency. Blue points indicate measured data with 1$\sigma$ error bars, and the red curve is a fit to $\mathcal{A}(\delta ) = \mathcal{C} \frac{\Omega^2}{\Omega_g^2}\sin^2\left(\frac{\Omega_gT}{2}\right)\left[\cos\left(\frac{\Omega_g T}{2}\right)\cos\left(\frac{\delta  \tau_{\text{free}} }{2}\right)-\frac{\delta}{\Omega_g}\sin\left(\frac{\Omega_g T}{2}\right)\sin\left(\frac{\delta  \tau_{\text{free}} }{2}\right)\right]^2 + A_0$, where $\delta = 2 \pi (f - f_{\text{res}})$ is the two-photon detuning,  $f_{\text{res}}$ is the resonant transition frequency, $\mathcal{C}$ is the Ramsey contrast, $\tau_{\text{free}}$ is the free evolution time between two pulses, $T$ is the pulse duration, $\Omega$ is the two-photon Rabi frequency, $\Omega_g = \sqrt{\Omega^2 + \delta^2 }$ is the generalized two-photon Rabi frequency, and $\mathcal{A}_0$ is the asymmetry offset. We incorporate a Gaussian distribution of molecular beam forward velocities which leads to a spread in $\tau_{\text{free}}$ and $T$.  The phase accumulation occurs during the Ramsey pulse as well so the Ramsey fringe period is $\tau \sim \tau_{\text{free}} + T$ in our case. The fitted value of $\tau_{\text{free}}$ and $T$ are $(15\pm 2)$ and $(10\pm 1)$ $\mu$s, respectively. (c) Level diagrams of the $N^{\prime\prime}=1$,  $\tilde{X}(010)$ state in ${}^{174}$YbOH at $\E=39.60$~V/cm and $\B=12.15$~G, highlighting the EDM-clock transition (ECT) that suppresses both electric and magnetic field sensitivities and nearby field-sensing transition (FST) used for field-sensing.} 
\label{fig:setup} 
\end{figure*}

In this work, we experimentally demonstrate a general method~\cite{Takahashi2023magicPRL} to engineer field-insensitive yet EDM-sensitive clock transitions in paramagnetic species, while simultaneously engineering transitions to sense external fields.  The concept of tuning parameters to magic values to suppress some unwanted effect has been successfully applied in a wide range of experiments~\cite{Derevianko2011_katori_lattice_clock, Ludlow2015, Kim2013, Ushijima2018,  Kotochigova2010Magic,Burchesky2021Rotational, Leung2023}, and is for example a critical ingredient for the high precision and accuracy of modern atomic clocks~\cite{Aeppli2024}.  We demonstrate that at particular ``magic'' values of applied electromagnetic fields, certain transitions in the YbOH molecule exhibit orders-of-magnitude suppression of electric \textit{and} magnetic field sensitivities
while maintaining a large (differential) internal effective electric field of $\Eeff\approx 22$~GV/cm. The observed suppression of the field sensitivities is measurement statistics-limited. This EDM-clock transition exhibits simultaneous zero-crossings of electric ($\bm{\E}$) and magnetic ($\bm{\B}$) field sensitivities at a particular electromagnetic field point (up to measurement precision), though the sensitivities are also considerably reduced over a broad range of electromagnetic field values.  We perform Ramsey measurements and show that coherence is maintained under large external electric and magnetic field fluctuations. 
We demonstrate a measurement protocol that utilizes both EDM-clock (field-insensitive) and sensing (field-sensitive) transitions and enables systematic error correction for EDM experiments.

This approach does not require any unique electronic structure, making it generically useful for diatomic or polyatomic molecules, either neutrals or ions, and for a broad range of CP-violating effects including eEDM, nuclear Schiff moments (NSM), and nuclear magnetic quadrupole moments (MQM)~\cite{Takahashi2023magicPRL}.  The basic idea is to identify pairs of states whose electric and magnetic sensitivities can be tuned with external fields to a point where they have vanishing differential Stark and Zeeman shifts, yet retain large differential CP-violation sensitivity.  The approach can use states that have a small difference in projection of total angular momentum ($M_F$), making state preparation and readout schemes straightforward.  These methods are particularly useful for species with heavy spinful nuclei, which are needed for NSM and MQM searches, and which typically have complex hyperfine structure.

Polyatomic molecules in parity doubled rovibrational states~\cite{Kozyryev2017PolyEDM}, such as the bending mode in linear triatomics like YbOH, have features useful for CP-violation searches. We find that for these molecules, the EDM-clock transitions of interest occur at relatively small applied electric fields $\lesssim$100~V/cm, adding a significant advantage to the protocol. 
In particular, we shall focus here on the eEDM-sensitive molecule $^{174}$YbOH~\cite{Kozyryev2017PolyEDM,Prasannaa2019,denis_enhancement_2019,Gaul2020}.  The energy shifts induced by the eEDM in a molecule under external fields are governed by the Hamiltonian~\cite{Flambaum2014}
\begin{multline}
H = -D\bm{\hat{n}}\cdot\bm{\E}-g\mu_B\bm{S}\cdot\bm{\B} + d_e\Eeff \bm{\hat{S}\cdot \bm{\hat{n}}}.
\end{multline}
$D$ is the molecular-frame dipole moment, $\bm{\hat{n}}$ is a unit vector along the molecular (symmetry) axis, $g$ denotes the magnetic $g$-factor, $\bm{\hat{S}}$ is the unit vector along the electron spin, $d_e$ is the electron EDM, and $\Eeff$ is the internal effective field experienced by the electron. The relative orientation of $\bm{\hat{S}}$ and $\bm{\hat{n}}$, defined as $P \equiv \bra \bm{\hat{S}\cdot \bm{\hat{n}}}\ket$, serves as a quantum-state-specific measure of the eEDM sensitivity. In this manuscript we refer to $\Eeff P $ as the ``eEDM sensitivity" since each state may have a different value of $P$.

\begin{figure}[t!]
    \includegraphics[width=0.9\linewidth]{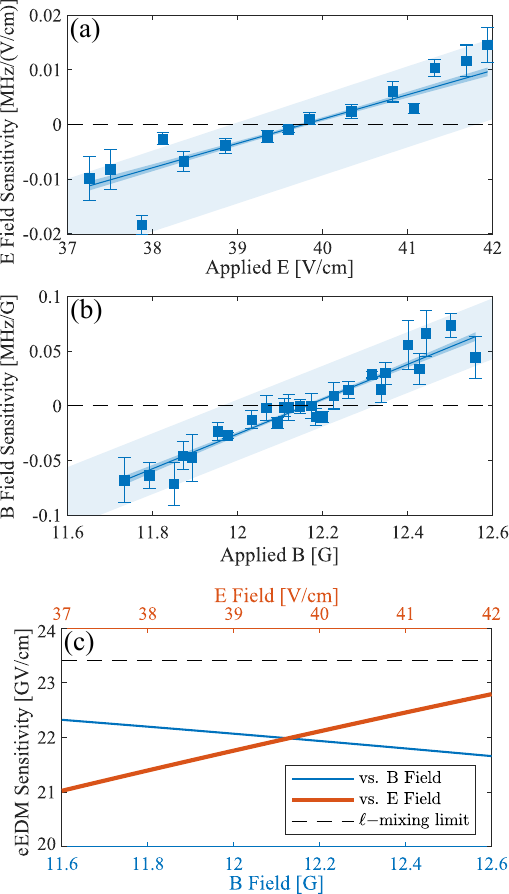}
\caption{\label{fig:field_sensitivity} Measured electric field sensitivity (a), measured magnetic field sensitivity (b), and calculated eEDM sensitivity (c) of the ECT as functions of the applied electric and magnetic fields. (a, b) Field sensitivities are expressed as the effective electric dipole moment and effective magnetic dipole moment. Blue square points represent measurements with their 1$\sigma$ errors, blue lines are linear fits, dark shaded regions indicate the fit uncertainties, and light shaded regions show the original theory uncertainty based on earlier spectroscopic measurements~\cite{Takahashi2023magicPRL}.  The applied magnetic field in (a) is 12.15 G, and the applied electric field in (b) is 39.60 V/cm. Linear fits yield 
$(\partial\Delta \mu_{\text{eff}}/\partial\B)= (1.6 \pm 0.1)\times 10^{-1}$ MHz/G$^2$ and $(\partial\Delta d_{\text{eff}}/\partial\E)= (4.4 \pm 0.3 )\times 10^{-3}$ MHz/(V/cm)$^2$, demonstrating that high suppression of \(\Delta d_{\mathrm{eff}}\) and \(\Delta \mu_{\mathrm{eff}}\) is maintained over a range of a few V/cm and \(\sim\)1 G, respectively. (c) The predicted eEDM sensitivity, based on molecular parameters provided in the Supplementary Material, is expressed in terms of $\Eeff \Delta P $ (GV/cm). The dashed black lines indicate the maximum eEDM sensitivity attainable in the limit of full mixing of the $\ell$ parity doublets.}
\label{fig:sensitivities}
\end{figure}

A schematic representation of the experiment is shown in Fig.~\ref{fig:setup}(a).  YbOH molecules are produced via laser ablation of pressed targets in a 4~K cryogenic buffer-gas cell and extracted as a cryogenic buffer gas beam (CBGB) with forward velocity $v\approx200$~m/s. After state preparation via 567~nm optical pumping into the ``science state'' $|\tilde X(010),N''=1,p=+\rangle$ manifold\footnote{Here $N$ denotes the molecular angular momentum not including spin, $p$ is the parity, and $(010)$ means that the molecule has one quanta of bending vibration.  This bending motion leads to parity doubling, making this state useful as the science state for EDM measurements~\cite{Kozyryev2017PolyEDM}.  $\tilde{X}$ denotes the ground electronic state.}, a Ramsey measurement on the transition of interest, denoted by the states $|0\ket$ and $|1\ket$, is performed via two-photon Raman pulses in a region of controllable electric and magnetic fields generated by indium-tin-oxide (ITO) coated glass plates and wire coils, respectively. There are additional few-meter-scale triaxial coils which null ambient magnetic fields to the mG level. Zeeman spectroscopy confirms that field uncertainties are $\sim1$~mG and therefore have negligible impact on our extraction of the field sensitivities. 

YbOH molecules in the beam experience two Ramsey pulses. The first $\pi/2$-pulse\footnote{For conceptual clarity we describe an ideal model with a perfect $\pi/2$-pulse, though our analysis includes imperfections in the pulses.} creates a superposition $|\psi\ket \propto |0\ket+|1\ket$. The state then evolves into $|\psi\ket \propto |0\ket+e^{i\varphi}|1\ket$ in the applied fields for a time $\tau\approx25~\mu$s over a distance of $\approx5$~mm. Here the phase $\varphi$ contains all physical phase accumulation, including the Stark and Zeeman shifts, the eEDM, and more~\cite{Baron2017}. The second $\pi/2$-pulse projects $\varphi$ onto the the populations in $|0\ket$ and $|1\ket$, yielding $|\psi\ket \propto \cos( \varphi/2)|0\ket+i\sin( \varphi/2)|1\ket$.
Finally, $\varphi$ is read out by laser-induced fluorescence with fast frequency switching~\cite{Kirilov2013} to form an asymmetry signal $\mathcal{A} = (N_{0}  - N_{1})/(N_{0} + N_{1})$ and suppress unwanted effects arising from beam yield fluctuations both within a single pulse and between different pulses. Note that imperfect optical pumping leaves populations in states close in energy to $|0\ket$ and $|1\ket$, and their fluorescence adds background ($\mathcal{A}_0$) and reduces the Ramsey contrast ($\mathcal{C}$).

Here we focus on a specific pair of transitions in ${}^{174}$YbOH -- an EDM-clock transition (ECT) whose sensitivities to both electric and magnetic fields are tuned to be simultaneously suppressed while maintaining high eEDM sensitivity, and a field-sensing transition (FST) with high sensitivity to electric and magnetic fields.  These transitions are shown in Fig.~\ref{fig:setup}(c) and are chosen for their experimental accessibility.  Note that there exist other engineered transitions with different sensitivities to fields and CP-violation effects~\cite{Takahashi2023magicPRL} (see Supplementary Material). We conduct two-photon spectroscopy to locate the line position of the ECTs at various electric and magnetic fields, and perform Ramsey measurements to measure their Stark and Zeeman shift to extract the field sensitivities.
The measured field sensitivities are shown in Fig.~\ref{fig:sensitivities}.

\begin{figure}[]
    \includegraphics[width=0.9\linewidth]{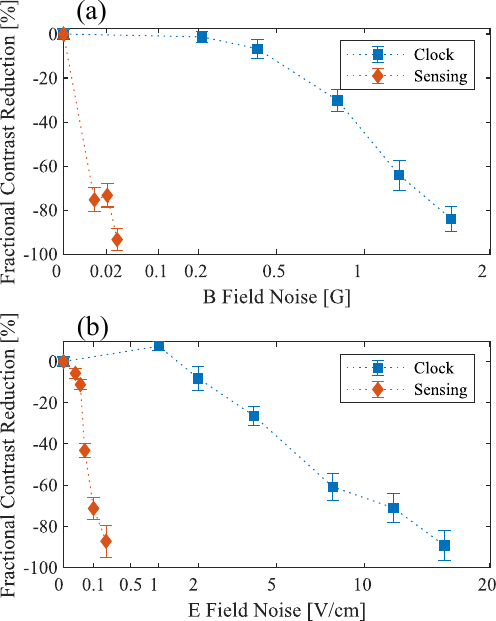}
\caption{\label{fig:noise_injection} Reduction in Ramsey fringe contrast as a function of electric and magnetic field noise amplitude. The noise has uniform distribution with amplitude defined as the difference between the minimum and maximum values. The $x-$axis scale is non-linear to more clearly show the behavior at low noise values. Data with 1$\sigma$ error bars in (a) were taken with zero applied electric field noise, and in (b) with zero applied magnetic field noise. The applied field noise distributions are centered at $\E=39.60$~V/cm and $\B=12.15$~G for the ECT, and $\E=20.00$~V/cm and $\B=0.30$~G for the FST. The field sensitivities of the FST around these fields are $\Delta d_{\text{eff}} = (0.120 \pm 0.001)$ MHz/(V/cm) $\sim 0.1 \Delta d_{\text{typ}}$ and $\Delta \mu_{\text{eff}}= (0.952 \pm 0.008)$ MHz/G $\sim \Delta \mu_{\text{typ}}$. The applied field noise distribution is discretized due to the finite resolution of the current and voltage supply for the magnetic coil and electric field plate, respectively. The applied fields sample 20-1300 discrete values for the ECT and 2–25 discrete values for the FST.}
\end{figure}

To quantify the sensitivity of a particular transition between two states $|0\ket$ and $|1\ket$ to $\bm{\B},\bm{\E}$, and $d_e$ we define the following differential quantities: the differential effective magnetic moment $\Delta \mu_{\text{eff}} \equiv \mu_{\text{eff}, 0}-\mu_{\text{eff}, 1}$ in MHz/G, the differential effective dipole moment $\Delta d_{\text{eff}}\equiv d_{\text{eff}, 0}-d_{\text{eff}, 1}$ in MHz/(V/cm), and the differential eEDM sensitivities $\Delta P\equiv(P_{0}-P_{1})$. 
For the \(\tilde{X}(010)\) state in YbOH, the maximum differential eEDM sensitivity obtainable in the limit of full mixing of the parity doublets is $\Delta P_{\text{max}}=|P_{N^{\prime\prime}=1, M_F=2, M_N \ell = 1} -P_{N^{\prime\prime}=1, M_F=2, M_N \ell = -1}| = 1$, where $M_N \ell$ indicates molecular orientation in the lab-frame axis.\footnote{Note that the angular momentum coupling in \( |\tilde{X}(010),N^{\prime\prime}=1\rangle \) results in the molecular polarization on the lab-frame $\hat{\bm{Z}}$ axis of $ \bra \bm{\hat{n}}\cdot\hat{\bm{Z}} \ket= \frac{M_N \ell}{N^{\prime\prime}(N^{\prime\prime}+1)}$, where \(\Delta P_{\text{max}}\) is smaller by a factor of two than in a $^3\Delta_1$  state. There are also local sensitivity maxima, which can result in \(\Delta P\) exceeding 1, as observed elsewhere~\cite{Petrov2022YbOHEField, Anderegg2023CaOHSpin}.}
Without employing engineered ECTs, typical transitions have electromagnetic sensitivities given by the molecular dipole moment $D=2.16$~Debye and Bohr magneton $\mu_B$~\cite{Jadbabaie2023YbOH010}. These parameters yield sensitivities of $\Delta d_{\text{typ}} \sim1$ MHz/(V/cm) and $\Delta \mu_{\text{typ}} \sim1.4$ MHz/G. 

The observed ECT field sensitivities exhibit zero crossings and are consistent with the predictions using the molecular parameters from previous optical spectroscopy to within the expected uncertainties~\cite{Jadbabaie2023YbOH010}. This indicates that ECTs can be located with only moderate uncertainty on spectroscopic parameters.  The observed field sensitivities are $\Delta d_{\text{eff}} = (-0.0009 \pm 0.0006)$ MHz/(V/cm) and $\Delta \mu_{\text{eff}}= (-0.0004 \pm 0.0055)$ MHz/G, or $|\Delta d_{\text{eff}}| < 0.0015$ MHz/(V/cm) and $|\Delta \mu_{\text{eff}}|< 0.006$ MHz/G. This corresponds to a suppression in electric and magnetic field sensitivities by at least a factor of 710 and 230 compared to $\Delta d_{\text{typ}}$ and $\Delta \mu_{\text{typ}}$ respectively. Note that the sensitivities are predicted to be even lower, so these limits correspond to our ability to measure them directly.

At the same time, the ECT maintains a large internal effective electric field of $\Eeff \Delta P \approx$22~GV/cm, or $ \Delta P /  \Delta P_{max} \gtrsim 93 \%$. This indicates that one can suppress the electromagnetic field sensitivities by orders-of-magnitude while maintaining high eEDM sensitivity. It is worth mentioning that our upper limit of magnetic sensitivity in the magic condition is on the same order of the $^3\Delta_1$ state in ThO~\cite{Kirilov2013} and HfF$^+$~\cite{Loh2013}.  

A key component of the Ramsey measurement is coherence. Suppressing decoherence enables longer coherence times and thus enhances the statistical sensitivity to the eEDM.
To demonstrate robustness against decoherence from electric and magnetic fields, we measure the contrast of a Ramsey fringe with applied field noise and compare the ECT to an FST with large field sensitivities. Note that the FST discussed here is different from the FST shown in Fig.~\ref{fig:setup} (c) and is chosen for its higher field sensitivities.  Applied field amplitudes are varied shot-to-shot with a uniform distribution. Ramsey contrast $\mathcal{C}$ is extracted through the measurement of asymmetry difference $\mathcal{A}(\delta =0) - \mathcal{A}(\delta = \frac{\pi}{\tau}) \propto \mathcal{C}$. The contrast of FSTs decays rapidly with field noise amplitudes of $\lesssim$0.1~V/cm and $\lesssim$0.02~G, whereas the ECTs tolerate noise amplitudes of $\lesssim$10 V/cm and $\lesssim$1 G, about two orders of magnitude larger, as shown in Fig.~\ref{fig:noise_injection} and consistent with predictions (see Supplemental Material.)

Our approach also provides the tools to sense and correct field imperfections which could give rise to eEDM-mimicking systematic errors -- a primary obstacle in experiments. In particular, background ``non-reversing'' electric and magnetic fields ($\E_{nr}$, $\B_{nr}$) which do not change when the applied fields are purposefully reversed are a common concern. Here, we show that alternating between ECT and FST measurements can sense the electric and magnetic environment directly.  Since the approach here uses the same rotational state in the molecule and the same lasers are used for the measurement, this approach can also be used to directly sense higher order effects which rely on the combination of multiple errors,  such as those observed with ThO~\cite{Baron2017}.  Since many nearby FSTs exist within the same electronic and vibrational state, alternating between ECT and FST is straightforward -- we simply adjust the two-photon frequency detuning. In our case we change the frequency of the acousto-optic modulator which generates the two-photon detuning by $\sim$200 kHz. Furthermore, because the transition can be connected via linearly polarized light in the presence of a magnetic field, one can reverse the eEDM shift by reversing only the electric or magnetic field, without changing the laser polarization.  Since the laser is polarized along the electric field direction, the laser does not need to be sent through the electric field plates, offering further technical advantages~\cite{Baron2017}.

\begin{figure}[]
    \includegraphics[width=0.9\linewidth]{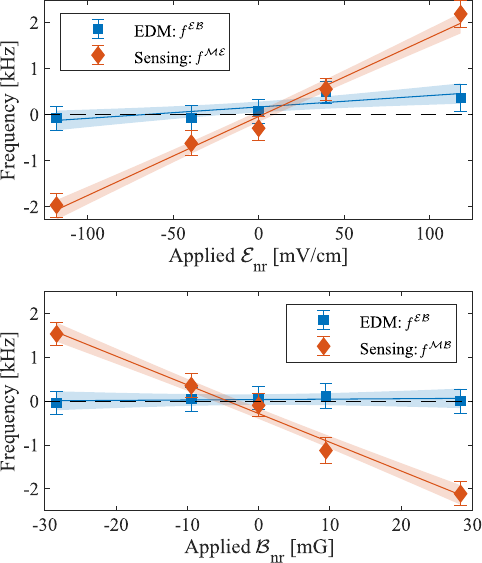}
\caption{\label{fig:systematics} \(f^{\M\E}\), \(f^{\M\B}\), and \(f^{\E\B}\) measured at various applied \(\E_{nr}\) and \(\B_{nr}\). Data in (a) were taken at applied \(\B_{nr} = 0\) G and in (b) at applied \(\E_{nr} = 0\) V/cm. Blue (square) and orange (diamond) points represent measurements with their 1\(\sigma\) errors, while the corresponding blue and orange solid lines are linear fits to the data.  The shaded regions indicate the uncertainty of the fit. The fit results for \(f^{\E\B}/\E_{nr}\) and \(f^{\E\B}/\B_{nr}\) are $(2.5 \pm 1.5)$ kHz/(V/cm) and $(1.0 \pm 6.4)$ kHz/G, which are smaller than that of \(f^{\M\E}/\E_{nr}\) and \(f^{\M\B}/\B_{nr}\), respectively.}
\label{fig:parity}
\end{figure}

To isolate the eEDM shift from other shifts, we measure the transition frequency $f$ under all combinations of three binary “switches”: the electric field up/down switch ($\hat{\E}$), the magnetic field up/down switch ($\hat{\B}$), and the ECT/FST (magic/non-magic) switch ($\hat{\M}$). The ECT and FST employed here are shown in Fig.~\ref{fig:setup} (c). We then take linear combinations of these eight frequencies to form “switch channels,” denoted by $f^s$,
so that each channel singles out effects correlated with the listed switches $s$~\cite{Baron2017}. For instance, $f^{\E\B}$ singles out the effects correlated with both $\hat{\E}$ and $\hat{\B}$ switches but not the $\hat{\M}$ switch, and contains both the genuine eEDM signal and potential eEDM-mimicking systematic effects. On the other hand, the switch channels $f^{\M\E}$ and $f^{\M\B}$ serve as sensitive probes of $\E_{nr}$ and $\B_{nr}$ with measured slopes $f^{\M\E}/ \E_{nr} = (17.2 \pm 1.5)$ kHz/(V/cm) and $f^{\M\B}/\B_{nr} = (-65.6 \pm 6.4)$ kHz/G, respectively (Fig. \ref{fig:systematics}). The uncertainties are computed from a statistical ensemble of 120 molecular pulses per data point.  This enables highly sensitive measurements of the electromagnetic fields directly, using the same science state, same lasers, same preparation and readout scheme, etc., thereby providing a very robust and easy-to-implement approach to field-sensing and co-magnetometry -- a feature challenging to realize with other approaches.

These measured fields can then be used to shim out field imperfections, or calculate and subtract systematic errors arising from them, analogous to the approach utilized in HfF$^+$~\cite{Caldwell2023_JILA_systematics}. Additionally, our scheme works for higher-order effects like field imperfections coupled to laser effects~\cite{Baron2017}, since the same lasers are used in the $\hat{\M}$ switch. There are also added benefits of high sensitivity to both electric \textit{and} magnetic fields, and the ability to directly measure them without relying on auxiliary measurements, in contrast to the scheme in ThO~\cite{Baron2017,Wu2020ThOQ}.  For example, offset fields can be deliberately applied to shim out non-reversing fields by minimizing shits in $f^{\M\E}$ and $f^{\M\B}$. Additionally, the false eEDM arising from a simultaneous non-reversing electric and magnetic field~\cite{Baron2017} $d_e^{\text{ false}}\propto \E_{nr}\B_{nr}$ can be determined from the $f^{\M\B}$ and $f^{\M\E}$ channels since $\E_{nr}\B_{nr}\propto f^{\M\B}f^{\M\E}$.    This shows that the switching between ECTs and FSTs serve as a robust protocol against systematic effects induced by uncontrolled electromagnetic fields. Other channels such as $f^{\E}, f^{\B}, f^{\M\E\B}$ also provide additional cross-checks of the eEDM signal and stray fields. Note that while $\E_{nr}$ and $\B_{nr}$ are important cases of systematic errors that we examine in detail in this work, other systematic errors can be introduced by additional imperfections, such as field gradients and transverse fields arising from misalignment between electric field, magnetic field, and laser polarization axis, all of which require careful consideration and measurement.  Importantly, a range of ECTs and FSTs with different sensitivities to different imperfections can be selected to diagnose and mitigate these systematic errors~\cite{Takahashi2023magicPRL}, in addition to auxiliary measurements and other standard approaches~\cite{Baron2017}.

In summary, we present the first experimental demonstration of engineered clock transitions that simultaneously suppress the sensitivities to electric and magnetic fields by orders-of-magnitude, enable sensing of the electromagnetic environment, and offer high sensitivity to fundamental symmetry violations.  The protocol is versatile and well-suited for a wide range of species, including those with large angular momentum commonly encountered in nuclear CP-violation searches, such as the ongoing effort in $^{173}$YbOH for a nuclear MQM search~\cite{Kozyryev2017PolyEDM,Pilgram2021YbOHOdd,Takahashi2023magicPRL}, and many more~\cite{Zhang2024_NSM_sensitivity, Stuntz2024, Sunaga_2024}.  Similar magnetically insensitive transitions have been demonstrated in rare-earth–ion-doped crystals~\cite{Fraval2004} and our scheme may be adopted for solid state dark matter and CP-violation searches~\cite{fan2024resultssearchaxionlikedark}.
The tunability of field sensitivities via application of external fields, combined with ultracold assembly or electromagnetic control techniques (electric and magnetic guiding, opto-electrical and ion trapping), could extend the protocol to non-laser-coolable species in symmetry-violation searches~\cite{Marc2023, Smiałkowski2021_silver_containing_molecules, Maximilian2025_Rempe_Zeppenfield_coherence_polyatomic}. Moreover, exploiting EDM-clock transitions for quantum-enhanced measurement schemes could boost CP-violating signals beyond the standard quantum limit while suppressing electromagnetic field sensitivities~\cite{Zhang2023Entangled}. Clock transitions in polarized molecules could also be potentially utilized for quantum simulation and computation that require long interaction times and long-range dipole–dipole interaction~\cite{Najafian_magic_for_qubitin_N2_2020, Prehn2021, Maximilian2025_Rempe_Zeppenfield_coherence_polyatomic}.
Our demonstration, together with these prospects, suggests that EDM-clock transitions can suppress decoherence and systematic errors across diverse platforms, holding promise for enhanced quantum sensing and metrology.

\begin{acknowledgments}

We thank John M. Doyle, Xing Fan, Zack Lasner, Kon Leung, Amar Vutha, Chandler Conn, and Phelan Yu for feedback on the manuscript. This work was supported by the Heising-Simons Foundation (2019-1193 and 2022-3361), the Rose Hills Foundation, and an NSF CAREER Award (PHY-1847550). Y.~T. was supported by an IQIM Eddleman Fellowship and the Masason Foundation. H.~D.~R. was supported by a Fulbright Scholarship. C.~Z. was supported by the Caltech David and Ellen Lee Postdoctoral Fellowship.  

\end{acknowledgments}

\medskip
\noindent\textbf{Author contributions:} Y.~T. developed the concept and, with N.~R.~H., conceived the project. Y.~T., H.~D.~R., Y.~Z., C.~Z. designed and built the experiment. Y.~T. and H.~D.~R. performed the measurements. Y.~T. and N.~R.~H. analyzed the data. Y.~T., H.~D.~R., C.~Z., A. J. performed spectroscopy to predict and identify molecular structure. Y.~T., A.~J., C.~Z. performed early investigations into the clock-based approach. Y.~T. and N.~R.~H. wrote the manuscript. N.~R.~H. supervised the project.

\medskip
\noindent\textbf{Corresponding author:} Correspondence should be addressed to Yuiki Takahashi (\href{mailto:yuiki@caltech.edu}{yuiki@caltech.edu})\\

\medskip
\noindent\textbf{Competing interests:} The authors declare no competing interests.

\bibliography{biblio,ref2}

\newpage
\onecolumngrid
\setcounter{section}{0}
\setcounter{equation}{0}
\setcounter{figure}{0}
\setcounter{table}{0}
\makeatletter

\newpage
\renewcommand{\thefigure}{S\arabic{figure}}
\renewcommand{\thetable}{S\arabic{table}}

\section*{Supplementary Material}

\FloatBarrier

\subsection{Methods}


In the experiment, YbOH molecules are produced by laser ablation of pressed powder targets at a $\sim$1 Hz repetition rate and subsequently thermalized by collisions with $\sim$4~K He buffer gas atoms in a cryogenic buffer gas cell. By resonantly driving the ${}^1S_0 \rightarrow {}^3P_1$ transition of ${}^{174}$Yb, the ${}^{174}$YbOH yield is increased by around an order of magnitude ~\cite{Jadbabaie2020}. YbOH is extracted through the cell aperture as a rotationally and translationally cold beam~\cite{Hutzler2012, Yuiki2021}.
The YbOH beam travels at $\sim200$~m/s through a room-temperature chamber where YbOH is optically pumped from $|\tilde{X}(000), N^{\prime\prime}=0, p=+\ket$ to $|\tilde{X}(010), N^{\prime\prime}=1, p=+\ket$ through the $|\tilde{A}(010), p=-\ket$ excited state using a laser at 567~nm~\cite{Arian_thesis}. There are no applied electric or magnetic fields in this region so the parity selection rules hold. 

The YbOH molecules then enter the ``science chamber'' located $\sim$112 cm downstream from the cell where the Ramsey measurement and laser-induced fluorescence (LIF) detection are performed. Inside the science chamber, two indium tin oxide (ITO) coated glass plates separated by 1~inch  apply a uniform electric field in the vertical direction, which defines the lab $\hat{\bm{Z}}$ axis.  Since the electric field applied in this work is small compared to the point where parity doublets are fully mixed~\cite{Jadbabaie2023YbOH010,Takahashi2023magicPRL}, a large portion of the population in $|N^{\prime\prime}=1, p=+\ket$ still remains in the states that are adiabatically connected before and after entering the science chamber. 

In the science chamber, YbOH molecules are illuminated with multiple lasers at 577~nm to perform Ramsey spin precession measurements.  Note that the lasers are always on, and the motion of the molecules through the fixed laser beams results in them experiencing temporal pulses. Here the transitions of interest consist of $|0\ket$ and $|1\ket$ in $|\tilde{X}(010),N^{\prime\prime}=1, p=+\ket$, roughly separated by $\sim$60 MHz. To be clear, depending on the measurement these may be different states. The states used for the $\hat{\M}$ switch (changing between the ECT and FST) demonstration are shown in Fig. 1 in the main text. The first laser optically pumps from  $|0\ket$ to $|1\ket$ through a $|\tilde{A}(010),p=-\ket$ excited state with $\hat{Z}$ polarization. Next the YbOH beam sees two Ramsey pulses. Note that we describe an ideal model with a perfect $\pi/2$-pulse for conceptual clarity, though our quantitative analysis uses a fit function that incorporates measured deviations from a perfect $\pi/2$-pulse.
The first is a $\pi/2$-pulse with a two-photon laser phase $\theta$ to create a superposition $|\psi(0)\ket\propto|0\ket+e^{i\theta}|1\ket$.  This state then evolves into $|\psi(\tau)\ket\propto|0\ket+e^{i(\varphi+\theta)}|1\ket$  under the applied fields for a time $\tau\approx$25 $\mu$s over a distance of $\approx$5~mm with $\sim 200$ m/s molecular beam forward velocity. Note that this distance is between the center of two Ramsey pulse lasers. The laser beam for the Ramsey pulses is elliptical, with the major axis oriented vertically. The measured second moment width of the Gaussian fit along the horizontal (molecular beam) axis is $2.88$ mm. The phase accumulation occurs during the Ramsey pulse as well so the Ramsey fringe period is $\tau \sim \tau_{\text{free}} + T$. The second Ramsey pulse projects the phase $\varphi$ onto the populations in $|0\ket$ and $|1\ket$.
Note that the second Ramsey pulse is applied by retro-reflecting the laser for the first Ramsey pulse, and therefore has a constant two-photon laser phase offset $\delta \theta$ from the first pulse, up to the fluctuation of the path length difference. The phase $\varphi$ contains all physical phase accumulation, including the Stark and Zeeman shifts, the eEDM, and more~\cite{Baron2017}. 
The Ramsey pulses are realized via two-photon Raman transitions with the driving laser detuned by approximately $0.5$~GHz from the optical resonance between $|\tilde{X}(010)\ket$ and $|\tilde{A}(010)\ket$. (For some field-sensing transitions, the detuning was set to 1~GHz.) A double-passed acousto-optic modulator (AOM) generates a single sideband with the carrier, and their beats address the ECTs and FSTs of interest. Finally, the last laser drives from $|0\ket$ or $|1\ket$ to $|\tilde{A}(010),p=-\ket$, and the resulting laser-induced fluorescence (LIF) from decays to $|\tilde{X}(010),p=+\ket$ is collected to give state populations from which $\varphi$ is inferred. All laser frequencies are locked to a wavemeter.


To normalize against molecular yield variations from pulse-to-pulse and the non-uniform temporal shape within each pulse we employ fast switching~\cite{Kirilov2013} of the detection laser frequency to alternate the LIF detection between $|0\ket$ and $|1\ket$ at 50 kHz. With this approach, we can construct the asymmetry $\mathcal{A} = (N_{0}  - N_{1})/(N_{0} + N_{1})$ within a single ablation pulse, which significantly reduces unwanted effects arising from pulse-to-pulse fluctuations and allows us to approach shot-noise-limited measurements. Here $N_{0}$ and $N_{1}$ are the population in $|0\ket$ and $|1\ket$, and typically $\sim$4500 and $\sim$2500 for the ECT, respectively. Note that, as mentioned in the main text, imperfect optical pumping leaves populations in states close in energy to $|0\ket$ and $|1\ket$, and the measured $N_{0}$ and $N_{1}$ include background signals from the population in these nearby states.  There are also background signals due to scattered laser light, which we subtract. They account for 50–70~$\%$ of the total fluorescence photon counts; therefore, our measurements are mostly limited by the combined photon shot noise of both the molecule fluorescence and the scattered laser light.   

\begin{figure}[h]
  \begin{minipage}[c]{0.57\textwidth}
    \includegraphics[width=0.9\textwidth]{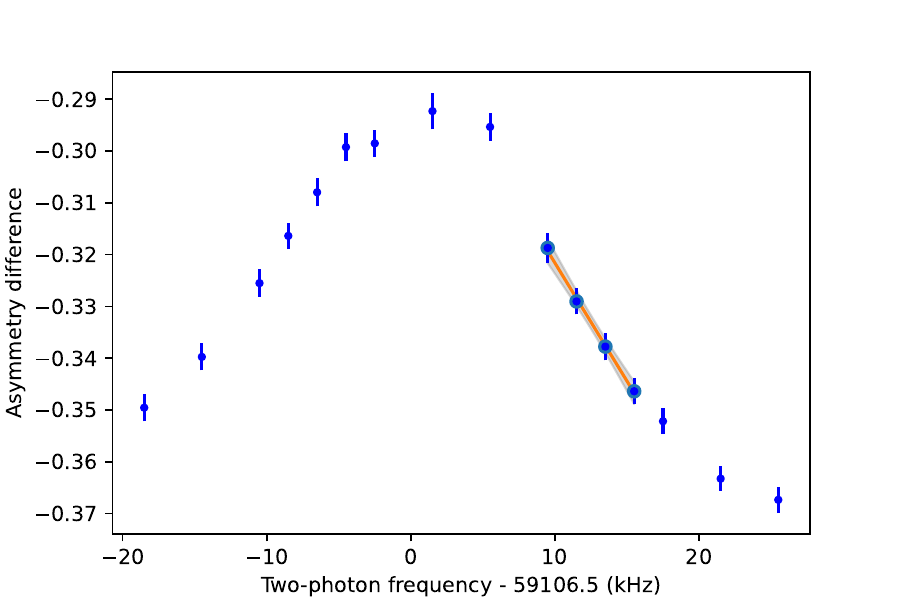}
  \end{minipage}\hfill
  \begin{minipage}[c]{0.4\textwidth}
    \caption{\label{sm_linear_portion} Ramsey fringe of the ECT as a function of two-photon frequency. The Blue points represent measurements with their 1\(\sigma\) errors, while the orange solid line is a linear fit to the linear portion of the fringe. The shaded regions indicate the uncertainty of the fit.}
  \end{minipage}
\end{figure}

As mentioned in the main text, we perform Ramsey measurements to conduct Stark and Zeeman spectroscopy and extract the field sensitivities. First, we set the two-photon frequency to lie within the linear portion of the Ramsey fringe, and fit the linear portion of the fringe with a linear function to obtain the slope $\mathcal{A}/f$, as can be seen from Fig.~\ref{sm_linear_portion}. Next, we vary the fields and record the asymmetries. As an example, we consider two asymmetry data points, $\mathcal{A}(\E_0, \B_0)$ and $\mathcal{A}(\E_1, \B_0)$, taken at $\E =\E_0, \B=\B_0$ and $\E =\E_1, \B=\B_0$, respectively, to extract the electric field sensitivity $\Delta d_{\text{eff}}$.  Note that during the measurements, we adjust the two-photon frequency accordingly so that two-photon frequency always lies within the linear portion of the Ramsey fringe.  We then take differences $\mathcal{A}(\E_0, \B_0) - \mathcal{A}(\E_1, \B_0)$ and divide it by the slope  $\mathcal{A}/f$ to convert it into a frequency difference $f(\E_0, \B_0) - f(\E_1, \B_0)$. The corresponding fit errors of the slope $\mathcal{A}/f$  are propagated. Finally, this frequency difference is divided by $\E_0 - \E_1$ to obtain the field sensitivity $\Delta d_{\text{eff}}(\frac{\E_0+\E_1}{2}, \B_0) = \frac{f(\E_0, \B_0) - f(\E_1, \B_0)}{\E_0 - \E_1}$. We report the field sensitivities at the midpoint ($\frac{\E_0+\E_1}{2}$) of two field points ($\E_0$ and $\E_1$), which is valid if the energy shift near the zero-crossing point is quadratic -- a behavior that is consistent with both our predictions and the observed data. Nonetheless, around the zero-crossing points, we have $|\E_0 - \E_1| \lesssim 0.5$ V/cm and $|\B_0 - \B_1| \lesssim 0.07$ G. ( $|\E_0 - \E_1|$ and $|\B_0 - \B_1|$ are different for different data points depending on their frequency shifts.) We repeat this measurement at various electric and magnetic fields to measure the field sensitivities $\Delta d_{\text{eff}}$ and $\Delta \mu_{\text{eff}}$. 

To suppress the effect of any possible drift of the overall asymmetry offset (e.g, due to the drift of the power balance between two frequency components in the fast switching of the detection laser), we alternate the two-photon frequency between red and blue side of the linear portion of the central Ramsey fringe. By taking the difference of the asymmetries at these two two-photon frequencies, we can cancel out the slow drift of the overall asymmetry offset. For the experimental data where we extract the bounds on the electric and magnetic field sensitivities at magic fields (i.e, the data at $\E=39.60$~V/cm and $\B=12.15$~G in Fig. 2 in the main text), we have taken additional precautions to measure the frequency shift accurately. First, instead of employing the switch mentioned above, we alternate the two electric and magnetic field values around the magic field values. This cancels out the systematic effect more directly. Additionally, though we have not observed a significant effect due to shot-by-shot variations of the molecular beam velocity (and thus Ramsey free evolution time) in the analysis time window within the molecular pulse (i.e, the variation of the asymmetry extracted from molecules arriving at the detection region within a specific time window), in order to suppress this effect, we integrate a large portion of the signal in each molecular pulse (4 ms out of $\sim$10 ms) even though this comes at the cost of reducing the frequency sensitivity. Finally, as explained in the next paragraph, we stabilize the two-photon laser power to suppress the effects of laser power drift to well below our measurement uncertainty.


For the two-photon transitions, the excitation pulses themselves can cause a frequency shift of the clock transition (probe light shift). To investigate this effect, we measure the transition frequency shift via Ramsey measurements at different two-photon laser powers. We change the laser power by $5\%$ with laser power stabilization. The stabilization is done by adjusting the RF amplitude fed into an AOM with an active feedback loop.  The measured energy shifts are $(0.8 \pm 0.8)$ kHz for the EDM-clock transition and $(1.2 \pm 0.5)$ kHz for the field-sensing transition. The laser power drift without power stabilization is typically less than $2\%$ over the experimentally relevant timescale of $\sim5$--$10$ minutes. This is the time needed to record a typical $\sim500$ shots at one electromagnetic field point, since the power can be measured and adjusted between field points. Thus, even without active feedback, any frequency shifts arising from laser power drift are smaller than the $1\sigma$ errors of this work. Nevertheless, we stabilize the two-photon laser power to make sure the laser power drift (and fluctuation) is well below $1\%$ and thus is negligible for the experimental data where we extract the bounds on the electric and magnetic field sensitivities at magic fields (i.e, the data at $\E=39.60$~V/cm and $\B=12.15$~G in Fig. 2 in the main text). Note that the $1\sigma$ errors of the switch channel frequency shifts ($f^s$) shown in Fig. 3 are $\sim 2\sqrt{2}$ times smaller than that of the transition frequency shifts $f\bigl( \hat{\mathcal{M}}, \hat{\mathcal{E}}|\mathcal{E}|, \hat{\mathcal{B}}|\mathcal{B}|\bigr)$ because of the way the switch channels are defined in the main text. Therefore, any possible shift due to the laser power drift is smaller than the $1\sigma$ errors shown in Fig. 3 as well.
The probe light AC Stark shift can also be categorized into switch channels and systematically characterized. This effect is well understood and characterized in atomic and molecular clocks, and there are methods to suppress and mitigate it via tailored pulse sequences~\cite{Santra2005, Zanon-Willette2006, Taichenachev2010, Yudin2010, Hobson2016} if needed, but also simply by measuring the relationship between laser power and the probe light shift~\cite{Leung2023}. Nonetheless, we can briefly estimate the magnitude of the several systematic effects that can arise from the combination of the non-reversing fields and probe light shifts. To simplify the estimation, we take the typical value of the two-photon Rabi frequency $\Omega_{eff} \sim  2 \pi \times 25$ kHz, one-photon detuning $\Delta \sim 2 \pi \times 1$ GHz, and one-photon Rabi frequency $\Omega \sim  2 \pi \times 10$ MHz. The non-reversing fields of $\E_{nr} \sim 1$  mV/cm and $\B_{nr} \sim 1$ mG can cause a shift of the one-photon detuning by $\delta \sim 1$ kHz, which correlates with the $\hat{\E}$ and $\hat{\B}$ switches, respectively. The probe light shift is roughly $\frac{\Omega^2}{4 \Delta}$ and since $\delta \ll \Delta$, the systematic shift in the probe light shift due to non-reversing fields are roughly $\frac{\Omega^2 \delta}{4 \Delta^2} \sim 1$ $\mu$Hz, which is already small. The two-photon Rabi frequency can also be changed by a similar, small amount.


A several meter-scale set of square coil pairs outside of the science chamber cancel background magnetic fields down to the $\sim$mG level in three axes. Another smaller pair of square coils applies a magnetic field of 0-20~G in the Ramsey region along the $\hat{\bm{Z}}$ axis.
We calibrated the magnetic field in the Ramsey measurement region by measuring the Zeeman shift of a magnetically sensitive transition in YbOH using two-photon spectroscopy. The uncertainty in this Zeeman shift measurement was found to be negligible for the final uncertainty of the obtained field sensitivities. Temporal fluctuations of the magnetic and electric fields are roughly 1 mG and 1 mV/cm (on the timescale of a second), respectively, and found to be negligible for the final results. 

Transverse fields can arise from misalignments between the electric field, magnetic field, and laser polarization axis. Similar to other EDM experiments~\cite{Baron2017}, these effects do not directly give rise to false EDMs but can couple in at higher orders. Therefore, these effects can also be categorized into switch channels and systematically characterized. To suppress the transverse fields, additional magnetic coils in the $\hat{\bm{X}}$ and $\hat{\bm{Y}}$ axis can be installed and magnetic field stabilization can be applied, as demonstrated elsewhere~\cite{Duan2022_B_field_active_feedback}. The $\hat{\bm{X}}$- and $\hat{\bm{Y}}$-axis-coils could also be useful for more precise control of the magnetic field orientation or for the investigation of systematic errors arising from transverse fields.


\subsection{Prediction of transition properties}

We previously performed high-resolution spectroscopy on the $\tilde{A}^2\Pi_{1/2}(000)-\tilde{X}^2\Sigma^+(010)$ band in YbOH~\cite{Jadbabaie2023YbOH010, Takahashi2023magicPRL} to extract the molecular parameters of the $\tilde{X}^2\Sigma^+(010)$ state. The uncertainties of the measured molecular parameters were $\lesssim$MHz, which is relatively high compared to the energy scale of Ramsey fringes in this work ($\lesssim$10 kHz).   Nevertheless, the theory prediction with the molecular parameters agrees with the properties of ECTs explored in this work within the uncertainties, as can be seen from Fig.~\ref{fig:field_sensitivity}. 
We can then manually adjust the molecular parameters within their spectroscopic uncertainties to better reproduce the observed field sensitivities. The adjusted molecular parameters used in this work are given in the Table~\ref{tab:sm}. Note that in the previous work, the hyperfine structure from the distant hydrogen nuclear spin was optically unresolved. Thus, we take the values from the previous two-photon spectroscopy work for the hydrogen hyperfine parameters (Fermi contact term $b_F$(H) and spin dipolar term $c$(H))~\cite{Arian_thesis}. 

\begingroup
\begin{table}[H]
\renewcommand{\arraystretch}{1.4}
 \centering
\begin{threeparttable}
 \caption{\label{tab:sm}
Spectroscopic parameters for the $\tilde{X}^2\Sigma^+(010)$ state of YbOH.}
\begin{ruledtabular}
\begin{tabularx}{1\textwidth}{llccc}

Parameter &	&  $^{174}$YbOH (previous work~\cite{Jadbabaie2023YbOH010}) &	 $^{174}$YbOH (adjusted in this work) \\ \hline 
$T_0$/cm$^{-1}$ & Origin energy &  319.90901(6) & 319.90901 \\
$B$/MHz & Rotation &  7,328.6(2) & 7,328.43$^b$ \\
$\gamma$/MHz & Spin-rotation & --88.7(9) & --87.7$^b$  \\
$\gamma_G$/MHz & Axial spin-rotation &  16(2) & 15.6$^b$  \\ 
$q_G$/MHz & $\ell$-doubling &  --12.0(2) & --12.2$^b$  \\
$p_G$/MHz & P-odd $\ell$-doubling &  --11(1) & --10.3$^b$ \\
$b_F$(H)/MHz & Fermi contact & 4.07(18)$^a$ & 4.07$^a$ \\
$c$(H)/MHz & Spin dipolar &  3.49(38)$^a$ & 3.49$^a$ \\
$D_{mol}$/Debye & Molecular dipole moment &  2.16(1) & 2.17$^b$ \\
$g_S$ & Effective electron g-factor &  2.07(2) & 2.07$^b$ \\

\end{tabularx}
\end{ruledtabular}
\begin{tablenotes}
\item{$^a$} These parameters are taken from the previous spectroscopy work~\cite{Arian_thesis}.
\item{$^b$} These parameters are manually adjusted in this work to better reproduce the observed field sensitivities.
\end{tablenotes}
\end{threeparttable}
\end{table}
\endgroup

\subsection{Other engineered transitions with different field sensitivities}

\begin{figure}[h]
    \includegraphics[width=0.9\textwidth]{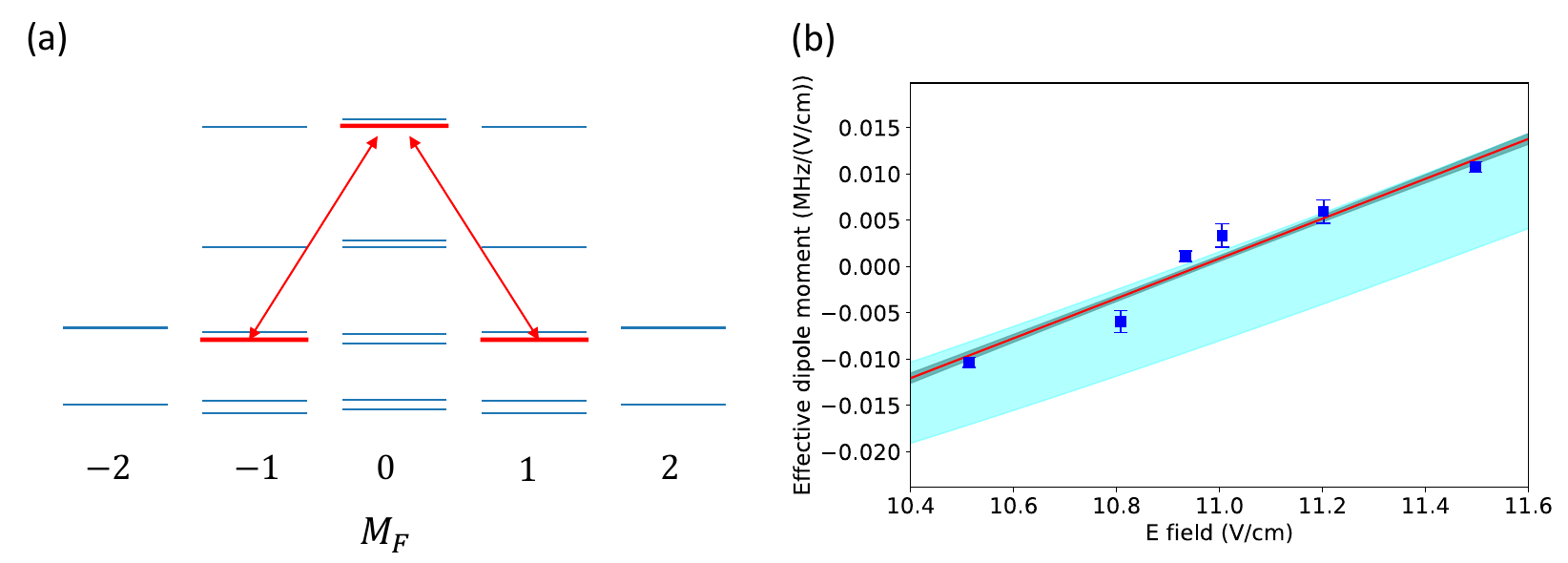}

 \caption{\label{sm_E_magic} (a) Level diagrams of the $N^{\prime\prime}=1$,  $\tilde{X}(010)$ state in ${}^{174}$YbOH at $\E=10.93$~V/cm and $\B=0$~G, showing an electrically insensitive, magnetically sensitive transition. (b) The electric field sensitivity as functions of the applied electric fields. We applied a $\sim$0.3~G magnetic field to lift the degeneracy between time-reversed pair for demonstration purposes. The electric field sensitivities are expressed in terms of the effective electric dipole moment. Blue square points represent measurements with their 1$\sigma$ errors, the red line is a linear fit, and the shaded gray region indicates the fit uncertainty.  The linear fit yields $(\partial\Delta d_{\text{eff}}/\partial\E)= (21.5 \pm 0.8 )\times 10^{-3}$ MHz/(V/cm)$^2$. The cyan shaded regions indicate the theory uncertainties, which are obtained from the measurement uncertainties of the previous work~\cite{Jadbabaie2023YbOH010} on the spectroscopic parameters. }
\end{figure}

Besides the ECT and FST explored in the main text, there are numerous transitions with different sensitivities to the electric field, magnetic field, and eEDM~\cite{Takahashi2023magicPRL}. We have indeed found other transitions experimentally, including ones with suppressed sensitivity to electric fields and high sensitivities to magnetic fields and the eEDM. Fig.~\ref{sm_E_magic} shows these transitions and their zero crossing of electric field sensitivity at a magic electric field value. This electrically insensitive, magnetically sensitive transition may offer an additional set of transitions useful for systematic effect searches. 
It is worth mentioning that the time-reversed pair of this transition exhibits identical electric field sensitivity but opposite signs of the eEDM sensitivity. This property is advantageous for systematic error rejection in a small magnetic field (like the case here), which is detailed in our previous publication~\cite{Takahashi2023magicPRL}.

\subsection{Comparison of Ramsey contrast with predictions}

\begin{figure}[h]
    \includegraphics[width=0.9\textwidth]{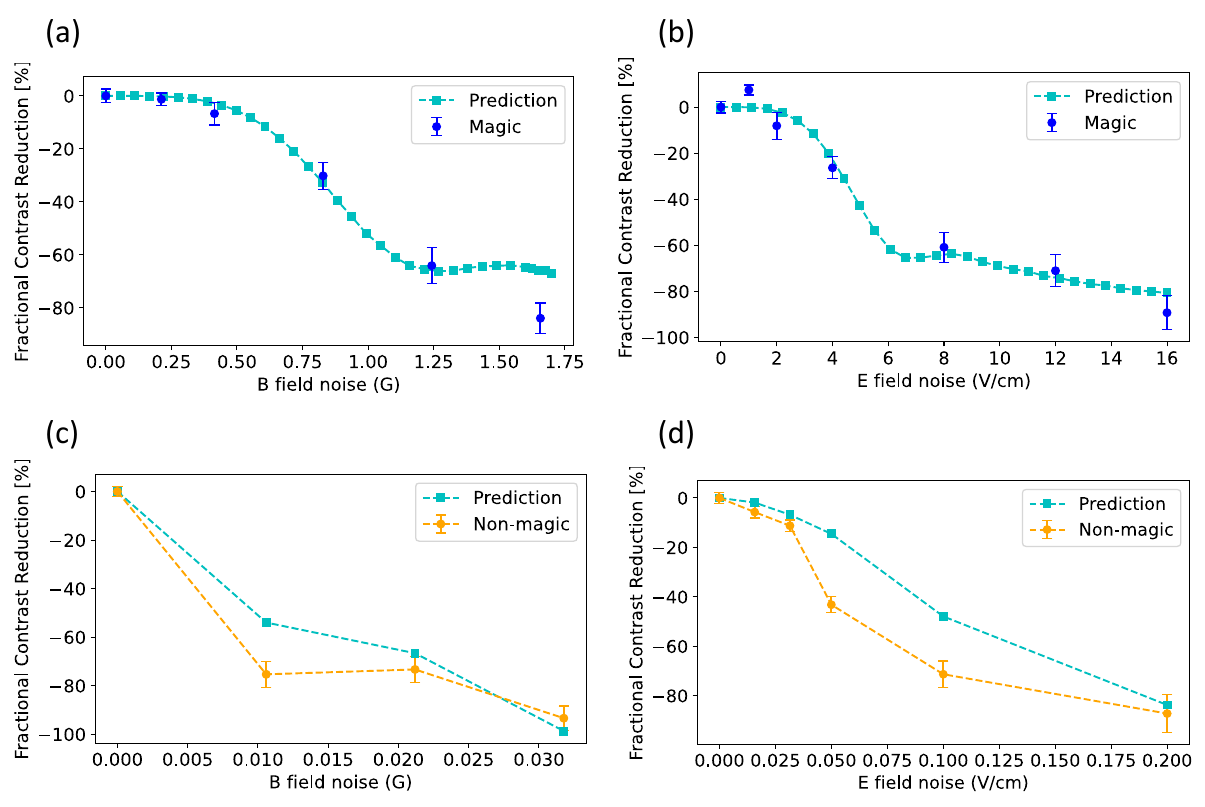}
 \caption{\label{sm_noise_injection_non_magic} Reduction in Ramsey fringe contrast as a function of electric and magnetic field noise amplitude. Data with 1$\sigma$ error bars in (a) and (c) were taken with zero applied electric field noise, and in (b) and (d) with zero applied magnetic field noise. The dashed cyan lines indicate the theory predictions. }
\end{figure}

Figure~\ref{sm_noise_injection_non_magic} shows the reduction of Ramsey fringe contrast as a function of electric and magnetic field noise amplitude. The noise has uniform distribution with amplitude defined as the difference between the minimum and maximum values. The applied field noise distribution is discretized due to the finite resolution of the current and voltage supply for the magnetic coil and electric field plate, respectively. For the EDM-clock transition (ECT), the applied fields sample 20-1300 discrete values, whereas for the field-sensing transition (FST), the small applied fields sample 2–25 discrete values. For the FST, the calculation explicitly includes the discretization of the applied field noise distribution. (For the ECT, since the number of the field samples is large, the calculation assumes a continuous noise distribution.) Note that they exhibit a reduced rate of decline at large noise amplitude (e.g, $\gtrsim$1.25 G and $\gtrsim$7 V/cm for the ECT) because strong fluctuations distort the original shape of the Ramsey fringe. 
For the FST, the data are plotted on an expanded scale for clarity.

\subsection{The switch channels}

The transition frequency $f\bigl( 
\hat{\mathcal{M}}, \hat{\mathcal{E}}|\mathcal{E}|, \hat{\mathcal{B}}|\mathcal{B}|\bigr)$ is measured under each experimental configuration defined by three binary switches: transition ($\hat{\mathcal{M}}$), electric field ($\hat{\E}$), and magnetic field ($\hat{\B}$).  The values of these switches is taken to be $\pm 1$ and corresponds to the two states of each parameter in which the experiment is operated: two directions of the electric field, two directions of the magnetic field, and two transitions.  Here $\hat{\E}=\mathrm{sign}(\bm{\E} \cdot \hat{\bm{Z}}), \hat{\B} = \mathrm{sign}(\bm{\B} \cdot \hat{\bm{Z}})$, and $\hat{\M} = +1~(-1)$ for the non-magic FST (magic ECT) transition. 

As mentioned in the main text, we form “switch channels,” denoted by $f^s$, where the superscript $s$ indicates the set of switches for which the quantity is odd under reversal.  This is a standard practice for eEDM experiments, and the details can be found in the literature~\cite{Baron2014}, but we summarize the important points and distinctions here.  These switch channels are defined as satisfying

\begin{equation}
\begin{split}
f(\hat{\mathcal{M}},\hat{\mathcal{E}}|\mathcal{E}|, \hat{\mathcal{B}}|\mathcal{B}|) 
&= f^{(0)}(|\mathcal{E}|, |\mathcal{B}|)
+ \hat{\mathcal{B}}\,\hat{f}^{\mathcal{B}}(|\mathcal{E}|, |\mathcal{B}|)
+ \hat{\mathcal{E}}\,\hat{f}^{\mathcal{E}}(|\mathcal{E}|, |\mathcal{B}|)
+ \hat{\mathcal{M}}\,\hat{f}^{\mathcal{M}}(|\mathcal{E}|, |\mathcal{B}|) \\
&+ \hat{\mathcal{M}}\hat{\mathcal{B}}\,\hat{f}^{\mathcal{M}\mathcal{B}}(|\mathcal{E}|, |\mathcal{B}|)
+ \hat{\mathcal{M}}\hat{\mathcal{E}}\,\hat{f}^{\mathcal{M}\mathcal{E}}(|\mathcal{E}|, |\mathcal{B}|)
+ \,\hat{\mathcal{E}}\hat{\mathcal{B}}\,\hat{f}^{\mathcal{E}\mathcal{B}}(|\mathcal{E}|, |\mathcal{B}|)
+ \,\hat{\mathcal{M}}\hat{\mathcal{E}}\hat{\mathcal{B}}\,\hat{f}^{\mathcal{M}\mathcal{E}\mathcal{B}}(|\mathcal{E}|, |\mathcal{B}|).
\end{split}
\end{equation}

This is a convenient parameterization since different phases can be categorized and distinguished by their switch channels.  The switch channels depend on $|\mathcal{E}|$ and $|\mathcal{B}|$, but not the switch states ($\hat{\M},\hat{\E},\hat{\B}$), and are calculated from the measurements by sums with either addition or subtraction depending on the switch channel, for example
\begin{eqnarray}
f^{(0)}
& = & \frac{1}{8} \sum_{\hat{\mathcal{M}}, \hat{\mathcal{E}}, \hat{\mathcal{B}}}
\,f\bigl( 
\hat{\mathcal{M}}, \hat{\mathcal{E}}|\mathcal{E}|, \hat{\mathcal{B}}|\mathcal{B}|\bigr)\\
f^{\mathcal{M}}
& = & \frac{1}{8} \sum_{\hat{\mathcal{M}}, \hat{\mathcal{E}}, \hat{\mathcal{B}}}
\hat{\mathcal{M}}\,f\bigl( 
\hat{\mathcal{M}}, \hat{\mathcal{E}}|\mathcal{E}|, \hat{\mathcal{B}}|\mathcal{B}|\bigr)\\
f^{\mathcal{E}\mathcal{B}}
& = & \frac{1}{8} \sum_{\hat{\mathcal{M}}, \hat{\mathcal{E}}, \hat{\mathcal{B}}}
\hat{\mathcal{E}}\,\hat{\mathcal{B}}\,f\bigl( 
\hat{\mathcal{M}}, \hat{\mathcal{E}}|\mathcal{E}|, \hat{\mathcal{B}}|\mathcal{B}|\bigr)\\
f^{\mathcal{M}\mathcal{E}\mathcal{B}}
& = & \frac{1}{8} \sum_{\hat{\mathcal{M}}, \hat{\mathcal{E}}, \hat{\mathcal{B}}}
\hat{\mathcal{M}}\,\hat{\mathcal{E}}\,\hat{\mathcal{B}}\,f\bigl( 
\hat{\mathcal{M}}, \hat{\mathcal{E}}|\mathcal{E}|, \hat{\mathcal{B}}|\mathcal{B}|\bigr)
\end{eqnarray}
with the rest defined similarly~\cite{Baron2017}. Note that here we choose to drop the explicit dependence of switch channels on $|\mathcal{E}|$ and $|\mathcal{B}|$. 
Standard error propagation of $f^s$ yields
\begin{equation}
\sigma_{f^s} = \frac{1}{8} \sqrt{\sum_{\hat{\mathcal{M}}, \hat{\mathcal{E}}, \hat{\mathcal{B}}}
\,\sigma_{f( 
\hat{\mathcal{M}}, \hat{\mathcal{E}}|\mathcal{E}|, \hat{\mathcal{B}}|\mathcal{B}|)}^2} := \sigma_{f}.
\end{equation}
Here, $\sigma_{f}$ is the same for all switch configurations $s$. Thus, the error of $f^s$ is the same for all switch channels.

The transition frequency of any transition depends on electric and magnetic fields due to Stark and Zeeman effects. Since we shall examine the transition frequency of ECTs where the first order linear Stark and Zeeman shifts vanish or become very small, we have to include the dependence of the transition frequency on electric and magnetic fields up to second order terms. It is worth noting that when we reverse the magnetic field we drive the transition with the different sign of the $M_F$ quantum number if we apply approximately the same frequency, such that the energies derived from the molecular, Stark, and Zeeman Hamiltonian are unchanged. We therefore express
\begin{eqnarray}
f(\E,\B) & = &  C_0 + C_{\E}|\E| + C_\B|\B| + C_{\E\B}|\E\B| + C_{\E\E}\E^2 + C_{\B\B}\B^2 \\
& = &  C_0 + C_{\E}\E\hat{\E} + C_\B\B\hat{\B} + C_{\E\B}\E\B\hat{\E}\hat{\B} + C_{\E\E}\E^2 + C_{\B\B}\B^2.
\end{eqnarray}

Here the transition frequency is expanded around some electric and magnetic fields ($\E_{0}$ and $\B_{0}$) that are in the vicinity of the electric and magnetic field magic points of the ECT. $C_0$ correspond to the transition frequency at $\E_{0}$ and $\B_{0}$ and all other coefficients are also given at $\E_{0}$ and $\B_{0}$. (Note that here we do not assume the existence of simultaneous zero-crossings of both the electric and magnetic field sensitivities at a particular electromagnetic field point.)
The absolute values are required to enforce the symmetry that in the absence of imperfections (or CP-violation) the transition energy shifts must be even under field reversals.  Notice again that when we change the sign of the magnetic field we are changing the sign of the $M_F$ quantum number if we address a transition with approximately the same frequency, so this symmetry with $\B$-reversal is enforced here unlike in most other eEDM experiments~\cite{Baron2017}.  The symmetry under $\E$-reversal arises from the usual $T$ and $P$ symmetry of the Hamiltonian.  These coefficients are determined by fitting to the calculated $\E$ and $\B$ dependence of the transitions around the predicted magic points of the ECT, and are shown in Tab.~\ref{tab:fit_EM}.

\begin{table}[H]
\centering

\end{table}

\begin{table}
  \begin{minipage}[c]{0.37\textwidth}
  \begin{tabular}{llrr}
Coefficient & Units & Magic & Non-magic \\ \hline
$C_{\E}$ & kHz/(V/cm) & 0 & $+45.04$ \\
$C_{\B}$ & kHz/G & 0 & $-197.20$ \\
$C_{\E\E}$ & kHz/(V/cm)$^2$ & $+2.59$ & $+2.61$ \\
$C_{\B\B}$ & kHz/G$^2$ & $+76.91$ & $+73.89$ \\
$C_{\E\B}$ & kHz/(G$\times$V/cm) & $-30.73$ & $-31.19$ \\
\end{tabular}
  \end{minipage}
  \hspace{4mm}
  \begin{minipage}[c]{0.55\textwidth}
\caption{Predicted coefficients in the electromagnetic sensitivities of the ECT and FST. 
Note that for these predictions we use spectroscopic parameters given in Tab.~\ref{tab:sm}, which have larger uncertainties from optical spectroscopy than the experimental energy scale of $\sim$few kHz in this work as explained above. Furthermore, the nuclear spin and rotational Zeeman terms are neglected.}
\label{tab:fit_EM}
  \end{minipage}
\end{table}

To include the effects of the state switch $\hat{\M}$, we can define  $\bar{C}_{\{\}}$ and $\delta \bar{C}_{\{\}}$ for all of the coefficients as the average and the half of the difference of the parameter $C_{\{\}}$ between the two transitions, respectively. For instance,
\begin{equation}
    \bar{C}_{\E} = (C_\E^\mathrm{FST} + C_\E^\mathrm{ECT})/2,\qquad
\delta \bar{C}_{\E} = (C_\E^\mathrm{FST} - C_\E^\mathrm{ECT})/2,
\end{equation}
and similarly for all other parameters.
Then, the transition frequency can be expressed as
\begin{equation}
\begin{split} \label{freq_eq}
f(\hat{\M}, \E,\B) & = (\bar{C}_0 + \hat{\M} \delta \bar{C}_{0}) + (\bar{C}_{\E} + \hat{\M} \delta \bar{C}_{\E}) \E\hat{\E} + (\bar{C}_\B + \hat{\M} \delta \bar{C}_{\B}) \B\hat{\B} + (\bar{C}_{\E\B} + \hat{\M} \delta \bar{C}_{\E\B})\E\B\hat{\E}\hat{\B} \\
& + (\bar{C}_{\E\E} + \hat{\M} \delta \bar{C}_{\E\E})\E^2 + (\bar{C}_{\B\B}+ \hat{\M} \delta \bar{C}_{\B\B})\B^2.
\end{split}
\end{equation}

Notice that the field sensitivities defined in the main text are the first derivative of the transition frequency with respect to $\E$ and $\B$,
\begin{eqnarray}
\Delta d_{\text{eff}} &  =  &  \frac{\partial}{\partial \E } f(\hat{\M}, \E,\B)  =  (\bar{C}_{\E} + \hat{\M} \delta \bar{C}_{\E}) 
 + (\bar{C}_{\E\B} + \hat{\M} \delta \bar{C}_{\E\B})\B + 2 (\bar{C}_{\E\E} + \hat{\M} \delta \bar{C}_{\E\E}) \E, \\
\Delta \mu_{\text{eff}} &  =  &  \frac{\partial}{\partial \B } f(\hat{\M}, \E,\B)  =  (\bar{C}_{\B} + \hat{\M} \delta \bar{C}_{\B}) 
 + (\bar{C}_{\E\B} + \hat{\M} \delta \bar{C}_{\E\B})\E + 2 (\bar{C}_{\B\B} + \hat{\M} \delta \bar{C}_{\B\B}) \B,
\end{eqnarray}
and thus their slopes (the second derivative of the transition frequency) are
\begin{eqnarray}
\frac{\partial}{\partial \E }  \Delta d_{\text{eff}} &  =  &  \frac{\partial^2}{\partial^2 \E } f(\hat{\M}, \E,\B)  =  
 2 (\bar{C}_{\E\E} + \hat{\M} \delta \bar{C}_{\E\E}), \\
\frac{\partial}{\partial \B } \Delta \mu_{\text{eff}} &  =  &  \frac{\partial ^2}{\partial ^2 \B } f(\hat{\M}, \E,\B)  =   2 (\bar{C}_{\B\B} + \hat{\M} \delta \bar{C}_{\B\B}). 
\end{eqnarray}

The electric and magnetic fields we apply in our experiment can be expressed as $(\E_{0}+\Delta\E)\hat{\E}$ and $(\B_{0}+\Delta\B)\hat{\B}$. Here $\Delta\E $ and $\Delta\B $ are the displacements of the fields from $\E_{0}$ and $\B_{0}$.
There are also non-reversing fields, which result in field magnitudes that change upon nominal reversal. Thus, the actual total fields we apply ($\E$ and $\B$) are the sum of ($\E_{0}\hat{\E}$ and $\B_{0}\hat{\B}$), displacements ($\Delta\E \hat{\E}$ and $\Delta\B \hat{\B}$) and non-reversing fields ($\E_{nr}$ and $\B_{nr}$), and we can therefore express the fields as
\begin{equation}\E = \E_{0}\hat{\E}  + \Delta\E\hat{\E} + \E_{nr}, \end{equation}
\begin{equation}\B = \B_{0}\hat{\B}  + \Delta\B\hat{\B} + \B_{nr}. \end{equation}
For clarity, we now focus only on systematic errors arising from non-reversing electric and magnetic fields, $\E_{nr}$ and $\B_{nr}$, which are some of the most common and important cases. As mentioned in the main text, other imperfections, such as field gradients or transverse components introduced by misalignment of the electric field, magnetic field, and laser-polarization axes, can also produce systematic errors. The significance of these additional imperfections depends on specific details of each experiment, and any future precision measurement must examine and suppress them carefully and thoroughly. In some cases, our protocol offers sensitive electric and magnetic field probes to assist these investigations~\cite{Caldwell2023_JILA_systematics}.  Importantly, a range of ECTs and FSTs with different sensitivities to different imperfections can be selected to diagnose and mitigate these systematic errors. 

Finally, we can include the eEDM shift $d_e\Eeff\hat{\E}\hat{\B}$.  Similar to the considerations needed for the Stark and Zeeman shifts, the eEDM shift here is manifestly $\E$- and $\B$- odd since reversal of $\B$ also effectively means reversal of $M_F$ and therefore the eEDM shift. We can understand this in the following way. First, we have $\bm{\hat{n}} = \hat{\E}\ \hat{\bm{Z}}$ since the sign$(H_{stark})$ of the state of interest does not change upon electric field reversal, where $H_{stark} = - \bm{\hat{D}} \cdot \hat{\E} - \bm{\hat{n}}\cdot \hat{\E}$~\cite{Lasner2019Thesis}. Note that in our case, we do not operate at fully polarized regime.  Second, we also have $\bm{\hat{S}} = \text{sign}(M_F)$ and $M_F$ is the projection of the total angular momentum along $\hat{\bm{Z}}$ axis. Again, the magnetic field reversal means $M_F$ reversal and therefore we have $\bm{\hat{S}} = \hat{\B}\ \hat{\bm{Z}}$. Finally,  the Hamiltonian for the eEDM term is $d_e\Eeff \bm{\hat{S}\cdot \bm{\hat{n}}} = d_e\Eeff \hat{\E} \hat{\B} $.

Using these parameterizations, and including the eEDM shift, we can build a table of the switch channels.

\begin{table}[H]
     \centering
    \begin{tabular}{c|l}
    Channel & Terms \\ \hline
$f^{(0)}$ & $\Delta\B\bar{C}_\B+\B_{nr}^2\bar{C}_{\B\B}+\Delta\B^2\bar{C}_{\B\B}+\Delta\E\bar{C}_\E+\Delta\B \Delta\E\bar{C}_{\E\B}+\Delta\E^2\bar{C}_{\E\E}+\E_{nr}^2\bar{C}_{\E\E}$ \\
$f^{\B}$ & $\B_{nr}\bar{C}_\B+2 \B_{nr} \Delta\B\bar{C}_{\B\B}+\B_{nr} \Delta\E\bar{C}_{\E\B}$ \\
$f^{\E}$ & $\E_{nr}\bar{C}_\E+\Delta\B \E_{nr}\bar{C}_{\E\B}+2 \Delta\E \E_{nr}\bar{C}_{\E\E}$ \\
$f^{\E\B}$ & $\B_{nr} \E_{nr} \bar{C}_{\E\B}+d_{{e}}\Eeff$ \\
$f^{\M}$ & $\Delta\B \delta C_\B+\B_{nr}^2 \delta C_{\B\B}+\Delta\B^2 \delta C_{\B\B}+\Delta\E \delta C_\E+\Delta\B \Delta\E \delta C_{\E\B}+\Delta\E^2 \delta C_{\E\E}+\E_{nr}^2 \delta C_{\E\E}$ \\
$f^{\M\B}$ & $\B_{nr} \delta C_\B+2 \B_{nr} \Delta\B \delta C_{\B\B}+\B_{nr} \Delta\E \delta C_{\E\B}$ \\
$f^{\M\E}$ & $\E_{nr} \delta C_\E+\Delta\B \E_{nr} \delta C_{\E\B}+2 \Delta\E \E_{nr} \delta C_{\E\E}$ \\
$f^{\M\E\B}$ & $\B_{nr} \E_{nr} \delta C_{\E\B}+ d_{{e}} \delta\Eeff$ \\
    \end{tabular}
     \caption{A table of switch channels in the general case.}
    \label{tab:parities_general}
\end{table}

Notice that $f^{\E} = \E_{nr} (\bar{C}_\E+\Delta\B \bar{C}_{\E\B}+2 \Delta\E \bar{C}_{\E\E}) = \E_{nr}\Delta \bar{d}_{\text{eff}}, f^{\B} = \B_{nr} (\bar{C}_\B+\Delta\E \bar{C}_{\E\B}+2 \Delta\B \bar{C}_{\B\B}) = \B_{nr} \Delta \bar{\mu}_{\text{eff}}$ where $\Delta \bar{d}_{\text{eff}}$ and $\Delta \bar{\mu}_{\text{eff}}$ are the average of the field sensitivities betweenthe  ECT and FST at $\E=\E_{0} + \Delta\E $ and $\B=\B_{0} + \Delta\B $,  and similarly for $f^{\M\E}$ and $f^{\M\B}$.  Because the ECT has significantly suppressed field sensitivities (i.e, $\Delta d_{\text{eff}}^\mathrm{ECT}, \Delta \mu_{\text{eff}}^\mathrm{ECT} \ll \Delta d_{\text{eff}}^\mathrm{FST}, \Delta \mu_{\text{eff}}^\mathrm{FST} $), we can write $\delta \Delta d_{\text{eff}}=\Delta \bar{d}_{\text{eff}}=\Delta d_{\text{eff}}^\mathrm{FST}/2:=d_{\text{eff}},\delta \Delta \mu_{\text{eff}}=\Delta \bar{\mu}_{\text{eff}}=\Delta \mu_{\text{eff}}^\mathrm{FST}/2:=\mu_{\text{eff}}$. 
We can thus further simplify the table:

\begin{table}[H]
     \centering
    \begin{tabular}{c|l}
    Channel & Terms \\ \hline
$f^{(0)}$ & $\Delta\B\bar{C}_\B+\B_{nr}^2\bar{C}_{\B\B}+\Delta\B^2\bar{C}_{\B\B}+\Delta\E\bar{C}_\E+\Delta\B \Delta\E\bar{C}_{\E\B}+\Delta\E^2\bar{C}_{\E\E}+\E_{nr}^2\bar{C}_{\E\E}$ \\
$f^{\B}$ & $\B_{nr} \mu_{\text{eff}}$ \\
$f^{\E}$ & $\E_{nr} d_{\text{eff}}$ \\
$f^{\E\B}$ & $\B_{nr} \E_{nr} \bar{C}_{\E\B}+d_{{e}} \bar{\E}_{eff} $ \\
$f^{\M}$ & $\Delta\B \delta C_\B+\B_{nr}^2 \delta C_{\B\B}+\Delta\B^2 \delta C_{\B\B}+\Delta\E \delta C_\E+\Delta\B \Delta\E \delta C_{\E\B}+\Delta\E^2 \delta C_{\E\E}+\E_{nr}^2 \delta C_{\E\E}$ \\
$f^{\M\B}$ & $\B_{nr}  \mu_{\text{eff}}$ \\
$f^{\M\E}$ & $\E_{nr} d_{\text{eff}}$ \\
$f^{\M\E\B}$ & $\B_{nr} \E_{nr} \delta \bar{C}_{\E\B} + d_{{e}}  \delta\bar{\E}_{eff}$ \\
    \end{tabular}
     \caption{A simplified table of switch channels.}
    \label{tab:parities_specific}
\end{table}

For the specific pair of transitions considered in the main text, the two transitions have the same $\Eeff$ to within 4\%, so $\delta\bar{\E}_{eff}\approx 0$. Therefore we regard $f^{\E\B}$ as the ``eEDM channel.'' Other transitions could be chosen with different relative eEDM sensitivity, providing further systematic checks of the eEDM channels.

As can be seen from Table \ref{tab:parities_specific}, $\E_{nr}$ and $\B_{nr}$ contribute to eEDM channel $f^{\E\B} =  \bar{C}_{\E\B} \E_{nr}\B_{nr} $ and also to $f^{\mathcal{M}\E} =  d_{\text{eff}} \E_{nr} $ and $f^{\mathcal{M}\B} =   \mu_{\text{eff}} \B_{nr} $. Thus, $f^{\mathcal{M}\E}$ and $f^{\mathcal{M}\B}$ serve as sensitive probes of $\E_{nr}$ and $\B_{nr}$, respectively, with much larger sensitivities than the eEDM channel $f^{\E\B}$. The direct non-reversing field systematic error can be expressed in more experimentally relevant units as $\bar{C}_{\E\B} \approx $30~$\mu$Hz/($\mu$G$\times$mV/cm), or $d_{e,false} \approx 6\times 10^{-30}$~e~cm/($\mu$G$\times$mV/cm). It is worth noting that $\bar{C}_{\E\B}$ is a parameter that depends only on molecular transition properties (but not experimental properties), and can be determined straightforwardly and independently from the slope measurements of switch channels such as  \(f^{\E\B}/\E_{nr}\) and \(f^{\E\B}/\B_{nr}\), for example, by direct Stark and Zeeman measurements. This allows us to distinguish between the direct non-reversing field systematic error and the systematic errors arising from the complicated higher-order effects like non-reversing fields coupling to laser effects if present.

Importantly, one can correct the false eEDM contribution by using the other channels, $Q =  \bar{C}_{\E\B} \E_{nr}\B_{nr} = \bar{C}_{\E\B} / ( d_{\text{eff}} \mu_{\text{eff}}) f^{\mathcal{M}\E} f^{\mathcal{M}\B} $.  Here $d_{{e}} \bar{\E}_{eff} 
 = f^{\E\B} - Q$.
When the field sensitivities and coefficients are determined with relative uncertainties of $\sigma_{ d_{\text{eff}}}/ d_{\text{eff}}, \sigma_{\mu_{\text{eff}}}/\mu_{\text{eff}}, \sigma_{\bar{C}_{\E\B}}/\bar{C}_{\E\B} < \sigma_{f}/f^{\mathcal{M}\E}, \sigma_{f}/f^{\mathcal{M}\B}$, error propagation yields 
\begin{eqnarray} 
\sigma_Q &=& Q\sqrt{\left(\frac{\sigma_{f}}{f^{\mathcal{M}\E}}\right)^2 + \left(\frac{\sigma_{f}}{f^{\mathcal{M}\B}}\right)^2 + \left(\frac{\sigma_{\bar{C}_{\E\B}}}{\bar{C}_{\E\B}}\right)^2 + \left(\frac{\sigma_{ d_{\text{eff}}}}{ d_{\text{eff}} }\right)^2 + \left(\frac{\sigma_{ \mu_{\text{eff}}}}{\mu_{\text{eff}}}\right)^2 + \frac{ 2 \sigma_{\E\B}}{f^{\mathcal{M}\E} f^{\mathcal{M}\B}} } \\
&<& Q\sqrt{\left(\frac{\sigma_{f}}{f^{\mathcal{M}\E}}\right)^2 + \left(\frac{\sigma_{f}}{f^{\mathcal{M}\B}}\right)^2 + \left(\frac{\sigma_{\bar{C}_{\E\B}}}{\bar{C}_{\E\B}}\right)^2 + \left(\frac{\sigma_{ d_{\text{eff}}}}{ d_{\text{eff}} }\right)^2 + \left(\frac{\sigma_{ \mu_{\text{eff}}}}{\mu_{\text{eff}}}\right)^2 + \frac{ 2 \sigma_{f}^2}{f^{\mathcal{M}\E} f^{\mathcal{M}\B}} }  \\
& \sim& Q\sqrt{\left(\frac{\sigma_{f}}{f^{\mathcal{M}\E}}\right)^2 + \left(\frac{\sigma_{f}}{f^{\mathcal{M}\B}}\right)^2 + \frac{ 2 \sigma_{f}^2}{f^{\mathcal{M}\E} f^{\mathcal{M}\B}}} \\
&=& \frac{\bar{C}_{\E\B} }{ d_{\text{eff}}  \mu_{\text{eff}}} \sqrt{\left(f^{\mathcal{M}\E}\right)^2 + \left(f^{\mathcal{M}\B}\right)^2 + 2 f^{\mathcal{M}\E} f^{\mathcal{M}\B}} \sigma_{f} \\
& \sim& 8 \times 10^{-6} [\frac{1}{\text{Hz}}] \sqrt{\left(f^{\mathcal{M}\E}\right)^2 + \left(f^{\mathcal{M}\B}\right)^2 + 2 f^{\mathcal{M}\E} f^{\mathcal{M}\B}} \sigma_{f} \\
& \ll& \sigma_{f}.
\end{eqnarray}
Thus, $\sigma_Q$ is suppressed compared to $\sigma_{f}$. Here, in eq. (19), we use $f^{\mathcal{M}\B}, f^{\mathcal{M}\E} \ll 100$ kHz. In eq. (14), $\sigma_{\E\B}$ is the covariance of $f^{\mathcal{M}\E}$ and $f^{\mathcal{M}\B}$, and follows

\begin{equation}
\begin{split}
\sigma_{\E\B} &= \frac{1}{64} \Bigl(  \sigma_{f(+1, +|\mathcal{E}|,+|\mathcal{B}|)}^2 -  \sigma_{f(+1, +|\mathcal{E}|,-|\mathcal{B}|)}^2 -  \sigma_{f(+1, -|\mathcal{E}|,+|\mathcal{B}|)}^2 +  \sigma_{f(+1, -|\mathcal{E}|,-|\mathcal{B}|)}^2 \\
&+  \sigma_{f(-1, +|\mathcal{E}|,+|\mathcal{B}|)}^2 -  \sigma_{f(-1, +|\mathcal{E}|,-|\mathcal{B}|)}^2 -  \sigma_{f(-1, -|\mathcal{E}|,+|\mathcal{B}|)}^2 +  \sigma_{f(-1, -|\mathcal{E}|,-|\mathcal{B}|)}^2 \Bigl) \\
&< \frac{1}{64} \sum_{\hat{\mathcal{M}}, \hat{\mathcal{E}}, \hat{\mathcal{B}}}\, \sigma_{f(\hat{\mathcal{M}}, \hat{\mathcal{E}}|\mathcal{E}|, \hat{\mathcal{B}}|\mathcal{B}|)}^2 = \sigma_{f}^2
\end{split}
\end{equation}
or simply, 
\begin{equation}
\begin{split}
\sigma_{\E\B } &= \rho_{\E\B}\sigma_{f^{\M\E}} \sigma_{f^{\M\B}}  = \rho_{\E\B}\sigma_{f}^2   \ll \sigma_{f}^2,
\end{split}
\end{equation}
where $\rho_{\E\B}$ is the correlation between $f^{\M\E}$ and $f^{\M\B}$  and satisfies $-1 \leq \rho \leq 1$.
We define covariance between $f^{\E\B}$ and $Q$ as $\sigma_{\E\B Q}$ and its magnitude satisfies the following,

\begin{equation}
\begin{split}
\sigma_{\E\B Q} &= \rho\sigma_{f^{\E\B}} \sigma_{Q}  = \rho\sigma_{f} \sigma_{Q}  \leq \sigma_{f} \sigma_{Q}   \ll \sigma_{f}^2,
\end{split}
\end{equation}
where $\rho$ is, again, the correlation between $f^{\E\B}$ and $Q$  and satisfies $-1 \leq \rho \leq 1$. Thus, the uncertainty of the term $d_{{e}} \bar{\E}_{eff}$ is dominated by the uncertainty of $f^{\E\B}$, $\sigma_{f}$.

\begin{figure*}
    \centering
    \includegraphics[width=1\textwidth]{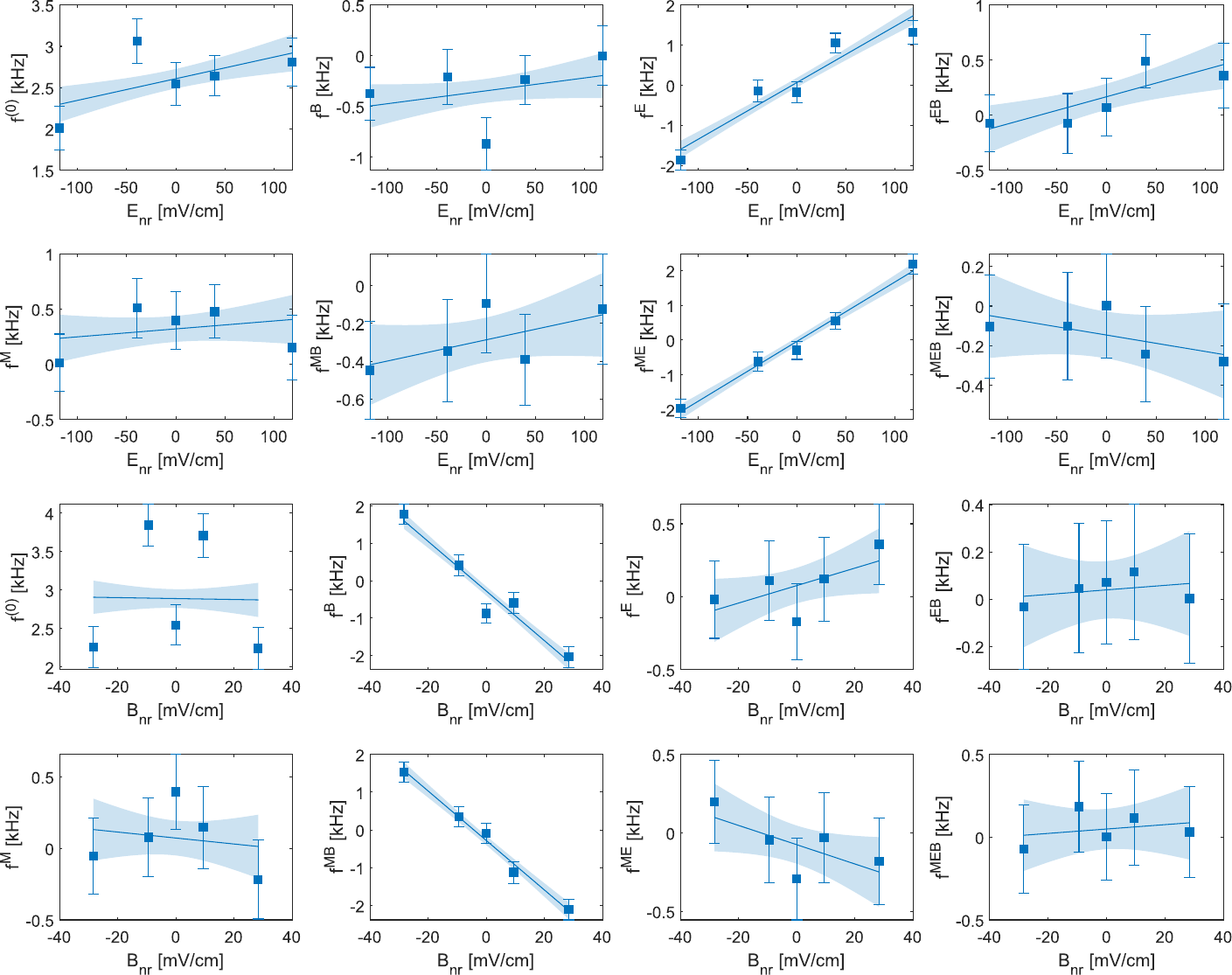}
    \caption{A plot of all switch parity channels vs. non-reversing fields.  Note that the $(0)$ channel is susceptible to overall drifts and is therefore expected to be noisy~\cite{Hutzler2014Thesis}.}
    \label{fig:allparities}
\end{figure*}

The high suppression of  $\bar{C}_{\E\B} / ( d_{\text{eff}}  \mu_{\text{eff}})$ relies on the high field sensitivities of the FST on the order of $\Delta d_{\text{typ}}$ and $\Delta \mu_{\text{typ}}$, and therefore is a general feature of FSTs. The result shows that switching between ECTs and FSTs serves as a robust protocol against systematic effects induced by non-reversing fields.
This systematic error correction protocol is very similar to that of the HfF$^+$ experiment~\cite{Caldwell2023_JILA_systematics}, which has effectively suppressed systematic errors arising from these effects. 
Note again that all the possible additional systematic shifts in $f^{\mathcal{M}\B}$, $f^{\mathcal{M}\E}$, and $f^{\E\B}$ must be thoroughly explored to measure and place limits on them, which can be assisted by our ``sensing channels'' such as $f^{\M\E}$ and $f^{\M\B}$, similarly to the HfF$^+$ experiment~\cite{Caldwell2023_JILA_systematics}. Again, a range of ECTs and FSTs with different sensitivities to different imperfections could further assist in investigation of the systematic errors.  
The FST here is chosen purely for its experimental accessibility; however, other FSTs with different eEDM and field sensitivities can be used if desired. 
Other channels such as $ f^{\E}, f^{\B}, f^{\mathcal{M}\E\B}$ also provide information on $\E_{nr}$, $\B_{nr}$, and the eEDM, offering an extra cross-check of the eEDM signal and its associated systematics, which can be seen from the Fig. \ref{fig:allparities}.  Here, $f^{\E}$ and $f^{\B}$ also show sharp linear responses to $\E_{nr}$ and $\B_{nr}$ with fitted slopes of $\Delta \bar{d}_{\text{eff}} = 14.1 \pm 1.5$ kHz/(V/cm) $\sim \delta \Delta d_{\text{eff}}$ and $\Delta \bar{\mu}_{\text{eff}} = -66.3 \pm 6.4$ kHz/G $\sim \delta \Delta \mu_{\text{eff}}$, respectively, thereby serving as another independent probe of $\E_{nr}$ and $\B_{nr}$.

\begin{table}
\begin{minipage}[c]{0.47\textwidth}
\begin{tabular}{c|ccc|ccc|}
    Channel & \multicolumn{3}{|c|}{Slope vs. $\E_{nr}$} & \multicolumn{3}{|c|}{Slope vs. $\B_{nr}$} \\ \hline
    & Pred & Meas. & Err. & Pred & Meas. & Err. \\
    $f^{(0)}$ & $0$ & $+2.6$ & $1.5$ & $0$ & $-0.61$ & $6.4$ \\ 
    $f^{\B}$ & $0$ & $+1.3$ & $1.5$ & $-99(39)$ & $-66$ & $6.4$ \\ 
    $f^{\E}$ & $+22(11)$ & $+14$ & $1.5$ & $0$ & $+6.0$ & $6.4$ \\ 
    $f^{\E\B}$ & $0$ & $+2.5$ & $1.5$ & $0$ & $+0.96$ & $6.4$ \\ 
    $f^{\M}$ & $0$ & $+0.72$ & $1.5$ & $0$ & $-2.1$ & $6.4$ \\ 
    $f^{\M\B}$ & $0$ & $+1.1$ & $1.5$ & $-99(39)$ & $-66$ & $6.4$ \\ 
    $f^{\M\E}$ & $+22(11)$ & $+17$ & $1.5$ & $0$ & $-6.1$ & $6.4$ \\ 
    $f^{\M\E\B}$ & $0$ & $-0.83$ & $1.5$ & $0$ & $+1.3$ & $6.4$ \\ 
    Units: & \multicolumn{3}{|c|}{Hz/(mV/cm)} & \multicolumn{3}{|c|}{Hz/mG}
    \end{tabular}
\end{minipage}
\hspace{4mm}
\begin{minipage}[c]{0.40\textwidth}
    \caption{A table of measured (Meas.) slopes with errors (Err.) and predictions (Pred.) for the different switch parity channels versus both $\E_{nr}$ and $\B_{nr}$.  The numbers in parentheses indicate prediction uncertainties from the spectroscopic parameter uncertainties.}
\end{minipage}
\end{table}

\end{document}